\documentclass[fleqn,usenatbib]{mnras}

\usepackage{newtxtext,newtxmath}

\usepackage[T1]{fontenc}
\usepackage{ae,aecompl}

\usepackage{graphicx}	
\usepackage{amsmath}	
\usepackage{rotating}
\usepackage{soul} 
\usepackage{xargs} 
\usepackage[pdftex,dvipsnames]{xcolor}  
\usepackage{color, colortbl}

\definecolor{shade}{gray}{0.9}
\title[Milky Way neutron stars from afar]{The Galactic neutron star population I - an extragalactic view of the Milky Way and the implications for fast radio bursts}

\author[A. A. Chrimes et al.]{
A. A. Chrimes,$^{1}$\thanks{E-mail: a.chrimes@astro.ru.nl}
A. J. Levan$^{1,2}$,
P. J. Groot$^{1,3,4}$,
J. D. Lyman$^{2}$
and G. Nelemans$^{1,5,6}$\\
$^{1}$Department of Astrophysics/IMAPP, Radboud University Nijmegen, P.O. Box 9010, 6500 GL Nijmegen, The Netherlands \\
$^{2}$Department of Physics, University of Warwick, Coventry CV4 7AL, UK \\
$^{3}$Inter-University Institute for Data Intensive Astronomy, Department of Astronomy, University of Cape Town, Private Bag X3, Rondebosch 7701, South Africa \\
$^{4}$South African Astronomical Observatory, P.O. Box 9, 7935 Observatory, South Africa \\
$^{5}$Institute for Astronomy, KU Leuven, Leuven, Belgium \\
$^{6}$SRON, Netherlands Institute for Space Research, Sorbonnelaan 2, NL-3584 CA Utrecht, The Netherlands \\
}

\date{Accepted XXX. Received YYY; in original form ZZZ}

\pubyear{2021}

\begin{document}
\label{firstpage}
\pagerange{\pageref{firstpage}--\pageref{lastpage}}
\maketitle

\begin{abstract}
A key tool astronomers have to investigate the nature of extragalactic transients is their position on their host galaxies. Galactocentric offsets, enclosed fluxes and the fraction of light statistic are widely used at different wavelengths to help infer the nature of transient progenitors. Motivated by the proposed link between magnetars and fast radio bursts (FRBs), we create a face-on image of the Milky Way using best estimates of its size, structure and colour. We place Galactic magnetars, pulsars, low mass and high mass X-ray binaries on this image, using the available distance information. Galactocentric offsets, enclosed fluxes and fraction of light distributions for these systems are compared to extragalactic transient samples. We find that FRBs follow the distributions for Galactic neutron stars closest, with 24 (75 per cent) of the Anderson-Darling tests we perform having a p-value greater than 0.05. This suggests that FRBs are located on their hosts in a manner consistent with Galactic neutron stars on the Milky Way's light, although we cannot determine which specific neutron star population is the best match. The Galactic distributions are consistent with other extragalactic transients much less often across the range of comparisons made, with type Ia SNe in second place, at only 33 per cent of tests exceeding 0.05. Overall, our results provide further support for FRB models invoking isolated young neutron stars, or binaries containing a neutron star.

\end{abstract}




\section{Introduction}
Several classes of astrophysical transients are associated with neutron stars, ranging from their formation in core-collapse supernovae to their mergers in short ${\gamma}$-ray bursts (GRBs). Magnetars are a subset of neutron stars \citep{1992ApJ...392L...9D}, distinguished by their extreme magnetic fields (10$^{14}$-10$^{15}$\,G), high radio/X-ray/${\gamma}$-ray luminosities despite spin-down luminosities comparable to pulsars \citep{1996ApJ...473..322T,1998Natur.393..235K}, and young characteristic ages of ${\sim}$10$^{3}$-10$^{5}$\,yrs \citep{2014ApJS..212....6O}. Magnetars are observed in the Milky Way and Magellenic clouds as soft ${\gamma}$ repeaters \citep[SGRs,][]{1979SvAL....5..343M,1979Natur.282..587M} and anomalous X-ray pulsars \citep[][]{1981Natur.293..202F}, detectable primarily through their high energy flares, and in some cases persistent emission \citep{2014ApJS..212....6O}. Giant magnetar flares, detectable at extragalactic distances, have been identified as an alternative origin for some short GRBs \citep{2005Natur.434.1098H,2006ApJ...652..507O,2008ApJ...681.1464O,2021arXiv210105104S,2021arXiv210105144B}. Magnetars have also been invoked as central engines in superluminous supernovae \citep{2010ApJ...717..245K,2010ApJ...719L.204W} and GRBs \citep{2008MNRAS.385.1455M,2011MNRAS.413.2031M}. More recently, they have become a promising explanation for the fast radio burst (FRB) phenomenon. 

This paper is partly motivated by the FRB-magnetar connection, following the detection of FRB-like bursts from the Galactic magnetar SGR\,1935$+$2154 \citep[hereafter SGR\,1935,][]{2020Natur.587...59B,2020Natur.587...54C}. Although the SGR\,1935 bursts are fainter than extragalactic FRBs, they are orders of magnitude brighter than other radio transients with comparable frequencies and millisecond durations \citep[e.g. pulsar pulses,][]{2018NatAs...2..865K}. Even before this detection, magnetar flaring was a leading theory among the dozens offered to explain FRBs \citep{2019PhR...821....1P}\footnote{\url{frbtheorycat.org/index.php/Main_Page}}. See also \citet{2020arXiv201210377C}, \citet{2020arXiv201210815B}, \citet{2021arXiv210104907X} and \citet{2020Natur.587...45Z} for reviews. Despite this association, there are still open questions around the relationship between the X-ray/${\gamma}$-ray and radio emission \citep{2016ApJ...826..226K,2020Natur.587...63L,2020ApJ...901L...7D,2021NatAs.tmp...31T,2021NatAs.tmp...30R,2021NatAs.tmp...48L,2021MNRAS.tmp..771B,2021arXiv210500685V,2021NatAs...5..408Y}. It is unclear if all FRBs repeat, or if there are two distinct populations of single and repeating events. Results from the first CHIME/FRB catalogue \citep{2021arXiv210604352T,2021arXiv210604356P} suggests that bursts from repeaters have larger temporal widths and narrow bandwidths, although the origin of this difference is currently uncertain. The nature of the weak periodicity/clustering seen in repeating FRB bursts also remains to be understood \citep{2020MNRAS.496.3390B,2020arXiv201208348P,2021arXiv210105836L}, although claims
of similar periodicity have now also been made for SGR\,1935 \citep{2021PASP..133g4202G}.

If we assume that FRBs are caused by magnetar flares, and that SGR\,1935 produces FRBs, then we must conclude that the Milky Way is an FRB host galaxy. If we also assume that all magnetars are capable of producing FRBs, then it follows that the distribution of FRBs on their host galaxies should match the distribution of magnetars on the Milky Way, at least for morphologically similar spiral galaxies. FRB sample size increases \citep[now $\sim$600 single FRBs and 20 repeaters,][]{2021arXiv210604352T} have lead to more host identifications \citep{2017ApJ...849..162E,2020ApJ...903..152H}\footnote{\url{frbhosts.org/}}, and consequently, population studies of these hosts have now begun. 

One way to study transient hosts is through their spectral energy distributions (SED), from which SED fitting can be used to infer the constituent stellar populations. Such work has found that FRB hosts span a range of masses and star formation rates, and correspondingly, morphological types. The mean stellar mass is around ${\sim}$10$^{10}$M$_{\odot}$, and star formation rates are typically moderate ${\sim}0.1-1$M$_{\odot}$yr$^{-1}$ \citep{2020ApJ...899L...6L,2020ApJ...895L..37B,2020ApJ...903..152H}, comparable to the Milky Way. These properties disfavour very massive stellar progenitors, such as those of superluminous supernovae (SLSNe) and long GRBs, but are consistent with regular core-collapse supernova (CCSN) hosts. To complicate matters further, \citet{2021arXiv210611993F} find that the host galaxy of FRB\,20201124A assembled $>90$\% of its stellar mass $>1$\,Gyr ago, perhaps favouring a long-delay time progenitor. Even more striking was the identification of a repeating FRB in an M81 globular cluster \citep{2021arXiv210511445K}, for which an old underlying population is overwhelmingly likely. Such scenarios are not straightforward to ascribe to magnetars given their young ages, and presumed origin in massive stars, although alternative models invoking accretion induced
collapse have been put forward \citep[e.g.][]{1992Natur.357..472U,2006MNRAS.368L...1L}.

The projected offset of transients from their host centre is another commonly used measure. At its most basic level, galaxies tend to have star formation rate gradients such that more star formation occurs towards the centre (neglecting metallicity effects and morphological differences). Furthermore, large offsets can provide insight into natal kicks, most prominently seen in short GRBs \citep{2010ApJ...708....9F,2011MNRAS.413.2004C,2014MNRAS.437.1495T}. \citet{2020ApJ...903..152H} and \citet{2020arXiv201211617M} measure the physical offsets ${\delta}r$ of FRBs from their hosts, and also the normalised offsets ${\delta}r/r_{e}$, where $r_{e}$ is the half-light radius. Using the normalised offset accounts for different radial intensity profiles (for example, the underlying stellar mass at a given offset can be different in two otherwise identical galaxies, if their physical size and compactness differ). \citet{2020ApJ...903..152H} and \citet{2020arXiv201211617M} find that CCSNe, SGRBs and type Ia SNe are the most consistent with FRB offsets. 

We can also measure the enclosed fraction of total galaxy flux within the galactocentric radius of the transient \citep{2006A&A...453...57J,2009MNRAS.399..559A,2015PASA...32...19A,2020MNRAS.492..848A}. A cumulative distribution of the fractional enclosed fluxes will produce a 1:1 relation if the transient traces the light. \citet{2020arXiv201211617M} calculate this distribution for 8 FRBs on their hosts in the IR, again finding consistency with a stellar mass tracing progenitor.

Another metric is the fraction of light \citep[F$_\mathrm{light}$,][]{2006Natur.441..463F}. This technique ranks host-associated pixels and normalises their cumulative distribution, such that the brightest pixel on a galaxy is assigned the value 1, representing the cumulative fraction of total host flux in regions of surface brightness `below' the pixel that contains the transient. Unbiased tracers of light sample F$_\mathrm{light}$ values uniformly, whereas biased datasets over/under sample from brighter/fainter pixels. Different wavelengths are used to probe different stellar populations: shorter, UV/blue bands trace young stars and thus star formation, IR/red bands better trace older stars and stellar mass. \citet{2020arXiv201211617M} compute F$_\mathrm{light}$ for 8 FRB hosts using UVIS and IR {\it Hubble Space Telescope}/WFC3 data. They compare to other transients, finding that type Ia SNe and SGRBs are the best match, but noting that the uncertainties are large. Nevertheless, this result suggests a stellar-mass tracing progenitor.

As magnetars have been suggested as both the source of FRBs, and as the central engines/remnants of SLSNe and GRBs, there have been several targeted radio searches at the locations of recently (i.e. in the last few decades) observed transients. Searches of this nature have yet to detect an FRB \citep{2017ApJ...841...14M,2020MNRAS.498.3863M,2020MNRAS.493.5170H,2021MNRAS.501..541P}, but deep optical searches have been able to rule out an association between FRBs and SLSNe/bright SNe Ia, GRB afterglows and tidal disruption events, on timescales of 1 day to 1 year post-burst \citep{2021arXiv210409727N}.

\begin{table*}
\centering 
\caption{The Galactic neutron star samples used in this work. Listed are the source reference, initial sample size, size following a luminosity cut (if applicable), and the final size, after restricting the sample to $y<8.3$\,kpc and removing objects in the Magellenic clouds (if previously included). } 
\label{tab:samples}
\begin{tabular}{lllllll}
\hline %
Population & Source & Init. size & Subset with & Luminosity cut & Size after & Final size after \\
 & & & $d$ estimate & & $L$ cut & $y<8.3$\.kpc cut \\
\hline %
Magnetars & McGill - \citet{2014ApJS..212....6O} & 31 & 26 & - & 26 & 20 \\
Pulsars & ANTF - \citet{2005AJ....129.1993M} & 2872 & 2822 & $>65$\,mJy\,kpc$^{2}$ & 229 & 127 \\
LMXBs & INTEGRAL - \citet{2020NewAR..8801536S} & 166 & 119 & - & 119 & 84 \\
HMXBs & INTEGRAL - \citet{2019NewAR..8601546K} & 64 & 57 & - & 57 & 36 \\
\hline %
\end{tabular}
\end{table*}

A final method to explore the nature of FRB progenitors is to compare their redshift distribution to the cosmic star formation rate (SFR) history. FRB population synthesis has again found conflicting results, with some work finding that their rate is consistent with the SFR \citep[and therefore magnetar production,][]{2020arXiv201206396G,2021arXiv210108005J}, and others claiming that they better trace stellar mass \citep{2020MNRAS.494..665L}

As can be seen from this overview of host populations, offsets (host level and resolved), F$_\mathrm{light}$, targeted searches for new FRB sources, and redshift distribution studies, the sole origin of FRBs in giant magnetar flares is still far from certain. Indeed, these results have led to conflicting interpretations over whether magnetars can explain all FRBs \citep{2020arXiv200913030B,2020ApJ...905L..30S}, and if so, what the progenitor channels should be \citep{2019ApJ...886..110M,2020ApJ...899L..27M}. Our comparison of magnetar (and other neutron star) positions in the Milky Way, versus FRBs locations within their hosts, is therefore a valuable alternative diagnostic.

The aims of this paper require the ability to map both the Milky Way itself, and the neutron star populations within in. Galactic magnetars have been studied in detail, in terms of their distances, activity, magnetic fields, ages and more \citep{2014ApJS..212....6O,2019MNRAS.487.1426B}; pulsar \citep{2005AJ....129.1993M,2017ApJ...835...29Y} and X-ray binary (XRB) catalogues are also maintained \citep{2019NewAR..8601546K,2020NewAR..8801536S}. Much of the previous literature has focused on modelling the Galaxy and neutron star populations in a statistical sense \citep[e.g.][]{2010A&A...510A..23S,2011ApJ...730....3S}, or if measured remnant positions are used, in terms of longitude, latitude or scale height distributions \citep[][]{1995ApJ...447L..33V,1996ApJ...473L..25W,2004MNRAS.354..355J,2005AJ....129.1993M,2014ApJS..212....6O,2017MNRAS.467..298R,2021arXiv210316973V}. However, there have been some attempts to measure correlations between neutron stars and Galactic structures in the plane of the Galaxy \citep[e.g. clustering between OB associations and high-mass X-ray binaries,][]{2012ApJ...744..108B,2013ApJ...764..185C}. Crucially, it is now possible to map spiral arms and star forming complexes throughout the Milky Way using masers \citep{2014ApJ...783..130R,2016ApJ...823...77R,2019ApJ...885..131R}. Doing this at other wavelengths, such as the optical, is difficult due to dust extinction, particularly further away from the Solar neighbourhood \citep[e.g.][]{2003A&A...397..133R,2014A&A...569A.125H,2016A&A...595A...1G}. There have also been studies looking at Milky Way global properties in the context of Milky Way analogues \citep[e.g.][]{2015ApJ...809...96L}. With precise localisations and distance estimates, we can now see how neutron stars would appear distributed on the Milky Way, if viewed externally. 

This paper is motivated by the magnetar-FRB connection, and as such there is a focus on magnetars and FRBs. However, studying how neutron stars are distributed throughout the Milky Way has a wider ranging relevance, for many classes of core-collapse and merger transients. This will be the focus of a follow-up paper in this series.

This paper is structured as follows. In Section \ref{sec:mappingNS} the Galactic magnetar, pulsar and X-ray binary samples are described. Section \ref{sec:mappingMW} describes how the basic components of the Milky Way are assembled in order to produce a face-on `image' of the Galaxy. The neutron star projected offsets, host normalised offsets and enclosed fluxes on the face-on disc of the Milky Way are discussed in Section \ref{sec:offsets}. Fraction of light measurements follow in Section \ref{sec:flight}. The arm offsets, host offsets, enclosed fluxes and fraction of light distributions are compared to samples of extragalactic transients in these sections. Section \ref{sec:discuss} presents a summary of the results and their implications, followed by concluding remarks in Section \ref{sec:conc}.

\section{Mapping Neutron Stars in the Milky Way}\label{sec:mappingNS}
Neutron stars can be identified observationally in several ways. Here, we use publicly available catalogues for four categories of neutron star/systems containing neutron stars. These are magnetars \citep[including SGRs and AXPs,][]{2014ApJS..212....6O}, pulsars \citep[including binaries and millisecond pulsars,][]{2005AJ....129.1993M}, and X-ray binaries \citep[high and low mass,][]{2019NewAR..8601546K,2020NewAR..8801536S}. These initial catalogues are reduced by selecting those sources with a distance estimate, and a luminosity cut is also applied to the pulsars, for increased uniformity of coverage across the Galaxy. 

The samples are then restricted to the half of the Galaxy this side of Galactic centre (GC). This is because the neutron star samples have biases against detection at large distances, particularly towards and beyond Galactic centre. Distance uncertainties also tend to be larger further away, for example, {\it Gaia} parallax-inferred distances (for XRBs) can only be estimated in the local 4-5\,kpc. Galactic centre is approximately 8\,kpc away \citep[e.g.][]{2019A&A...625L..10G}, but throughout this paper we adopt a $y$-coordinate cut of $<8.3$\,kpc (we define the $y$-axis as parallel to the Sun-Galactic centre line in a face-on Galactic coordinate system, with the Sun at 0 and GC at 8.3\,kpc). This cut approximately halves the galaxy at Galactic centre, but ensures that the centrally located magnetar SGR\,1745$-$2900 is not excluded from any distributions. This would be unrepresentative given that we know of at least one magnetar in the central ${\sim}100$\,pc of the Galaxy. Because we expect that Galactic populations on our side, versus the other side, of Galactic centre should be similar, these half-Galaxy samples should be broadly representative of the Galaxy as a whole. Any objects in the Magellenic clouds are also removed in this final cut. 

The input catalogues, cuts and final samples used are summarised in Table \ref{tab:samples}. We now describe each neutron star data set in turn.

\subsection{Magnetars}
We use the McGill magnetar catalogue of \citet{2014ApJS..212....6O}. The distances for each source are the most recent values in the literature, and the reasoning for each distance is listed in Table 7 of \citet{2014ApJS..212....6O}. These include estimating the distance to an associated SNR, using the tip of the red giant branch in associated clusters, and X-ray inferred column densities. Heliocentric distances uncertainties and are given by the upper and lower values listed in the McGill catalogue (and distance references therein). Where an uncertainty has not been quantified, a 15 per cent error on the distance is assumed, the mean of the uncertainties that are available.

The distance for SGR\,1935 is highly uncertain, with estimates from dispersion measures, dust extinction, a potentially associated SNR, and nearby molecular clouds \citep[see][for an overview]{2021MNRAS.tmp..771B}. We adopt the \citet{2021MNRAS.tmp..771B} best distance estimate of 6.5\,kpc, with a lower limit from their dispersion measure/dust analysis of 1.2\,kpc, and an upper limit of 9.5\,kpc \,the value assumed by \citet{2020Natur.587...59B}.

Although the lack of a strong local bias in the magnetar sample suggests that selection effects are not significant \citep{2014ApJS..212....6O}, there is a dearth of magnetars on the far side of the Galactic centre. This is despite the fact that ${\gamma}$-ray observatories have sufficient sensitivity to detect giant magnetar flares at extragalactic distances \citep{2021arXiv210105144B}. A similar effect has been noted before in the X-ray binary population \citep{2004MNRAS.354..355J}, although in that case, detections are in softer X-rays which are more liable to HI absorption and source confusion effects that preferentially occur along sightlines towards the Galactic centre. 

There are three possibilities: that many of the magnetars have mis-assigned distances, that the sample really is incomplete, or that the effect is real and there fewer magnetars on the far side of the Galaxy (unlikely, given that spiral galaxies tend not to have large scale asymmetries of this nature). \citet{2019MNRAS.487.1426B} discuss the Milky Way magnetar sample, explaining that calculating completeness is non trivial when SGRs are detected in different modes (quiescence versus flaring) with different observatories. They conclude that the missing fraction is around 0.3 for currently active magnetars. Similarly, \citet{2015MNRAS.454..615G} find that the Galactic X-ray pulsar population is complete above an absorbed flux of 3$\times$10$^{-12}$\,erg\,s$^{-1}$\,cm$^{-2}$. The sample used is therefore unlikely to be seriously incomplete for currently active sources this side of Galactic centre. The magnetars in the final sample are listed in Table \ref{tab:info} of appendix \ref{apx:A}.

\begin{figure}
	\includegraphics[width=\columnwidth]{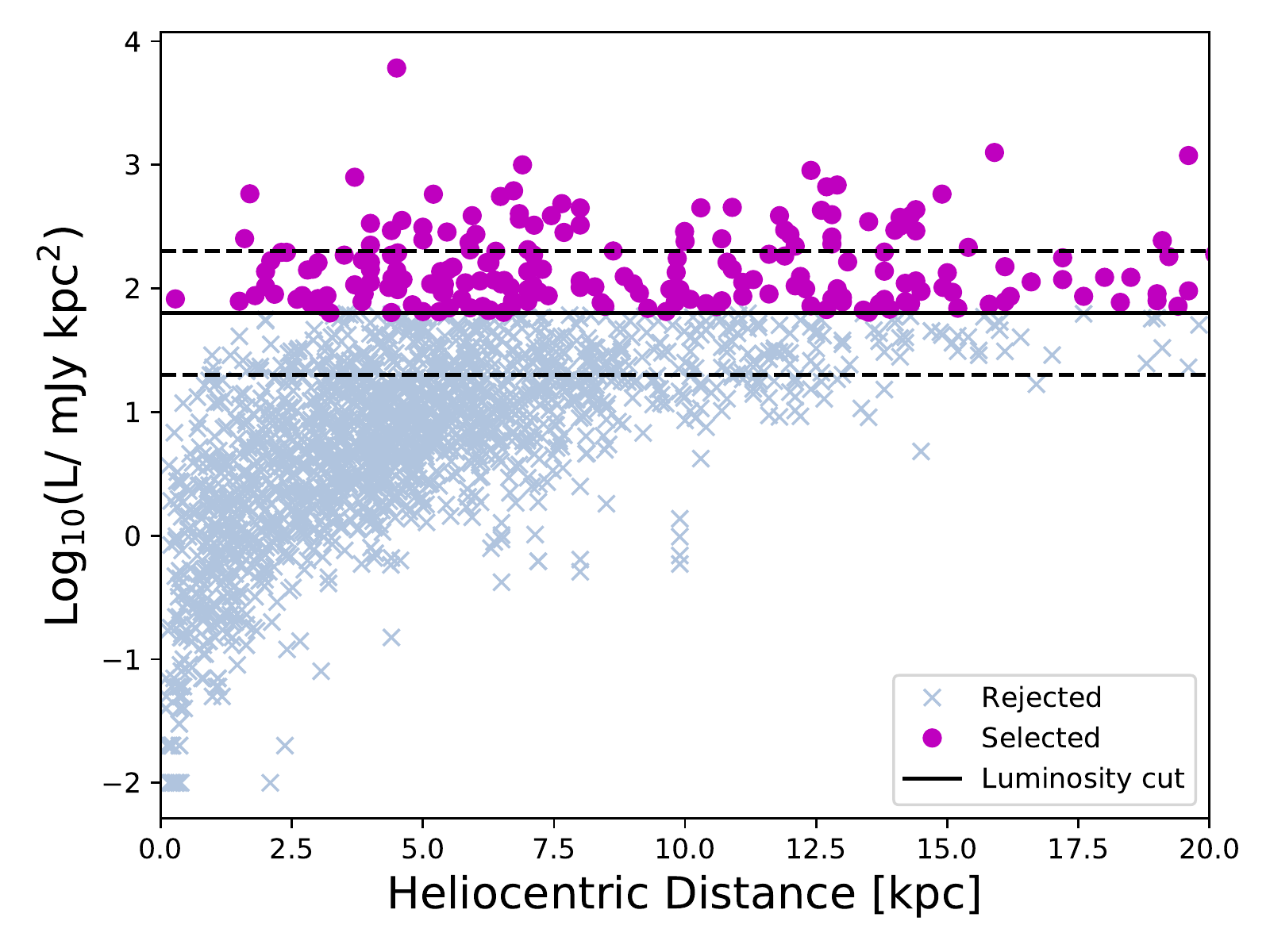}
    \caption{ATNF catalogue pulsar 1.4\,GHz luminosities versus the best distance estimates. A cut is applied at 65\,mJy\,kpc$^{2}$, which removes much of the strong local disc bias in the sample and gives (approximately, with the exception of the Galactic centre) equal coverage across the Galaxy. Two alternative cuts ($\pm$0.5 dex) are shown by dashed lines, these are later used to demonstrate the impact of different choices. Selecting the most luminous pulsars slightly shifts the sample towards younger sources - in log$_{10}$(age/yrs), the full sample has a median and standard deviation of 6.8$\pm$1.4, compared to 6.5$\pm$0.9 for the bright pulsars.}
    \label{fig:pulsar_lum}
\end{figure}

\subsection{Pulsars}
For pulsars, we use the ATNF catalogue \citep[version 1.64,][]{2005AJ....129.1993M,2016yCat....1.2034M}. Pulsars are far more numerous and have substantially weaker magnetic fields than active magnetars, typically in the range 10$^{11}$-10$^{13}$\,G. The catalogue provides best estimate distances, inferred from parallaxes, HI absorption, globular cluster associations, nebular lines and stellar companions. Otherwise, the dispersion measure is used to infer a distance, in this case using the \citet{2017ApJ...835...29Y} electron density model. We note that the majority of the sample are reliant on this model for a distance estimate, and will therefore be concentrated into the high electron-density spiral arms. This is less of an issue in the local disc, where other distance measurements are more prevalent.

Incompleteness is likely higher in this sample than the magnetars, as shown in Fig. \ref{fig:pulsar_lum}. However, pulsars are sufficiently numerous that a cut can be placed at high luminosity, such that a large region is covered in a less biased way, with the exception of the known dearth of pulsars in the Galactic centre. This detection bias along central sightlines arises due to the high dispersion measure here \citep{2017MNRAS.471..730R}. We will later demonstrate the effect this has by comparing pulsar fraction of light distributions that include and exclude the Galactic bulge region.

The chosen cut is 65 mJy\,kpc$^{2}$. This is the lowest cut that yields approximately constant numbers across the whole Galaxy (specifically, it yields numbers either side of the GC heliocentric distance that are consistent within Poisson uncertainties). To make the numbers exactly equal across both halves of the Galaxy requires a harsh cut, reducing the total to only a handful of sources. Our chosen cutoff is a compromise between sample size and completeness. We later vary this threshold to determine the impact on our fraction of light results. The $y<8.3$\,kpc half-Galaxy cut is applied {\it after} the luminosity threshold is applied to the whole catalogue.

\subsection{Low-mass X-ray binaries}
XRBs are systems containing an accreting black hole or neutron star; low mass XRBs (LMXBs) have a donor star masses that are typically ${\lesssim}2.5$M$_{\odot}$. Although still subject to luminosity-distance completeness issues, XRBs benefit from having a maximum theoretical luminosity (the Eddington luminosity) and therefore distances can be approximated from X-ray observations alone, in the absence of other indicators \citep{2004MNRAS.354..355J}. We use the INTEGRAL \citep{2003A&A...411L...1W} sample of \citet{2020NewAR..8801536S}, consisting of LMXBs detected by the hard X-ray sensitive telescope IBIS. Some distances in the catalogue are estimated from {\it Gaia} parallaxes of the optical counterpart, which may be introducing a local bias in the sample with measured distances.

\citet{2017MNRAS.470..512K} show the distance contours at which X-ray sources of given luminosities can be detected in the INTEGRAL survey. The majority of the Galaxy is covered with high completeness down to hard X-ray luminosities of 2${\times}$10$^{35}$\,ergs$^{-1}$, which is the lower end of the LMXB luminosity distribution. The sensitivity map provided by \citet{2017MNRAS.470..512K} shows that coverage is not uniform, as total survey time varies with Galactic longitude. However, the contours can be approximated as circular from a location that is ${\sim}3-5$\,kpc closer to the Galactic centre than the Sun \citep[see figure 2 of][]{2017MNRAS.470..512K}. Since the 2${\times}$10$^{35}$\,ergs$^{-1}$ limit is ${\sim}$20\,kpc away from this point in all directions, only the outskirts of the Galaxy are affected by survey biases, which should be of minimal impact. As the accretor type (i.e. neutron star or black hole) is often ambiguous, we do not attempt to separate them. The XRB populations in this paper therefore include a small contribution from black hole systems.

\subsection{High-mass X-ray binaries}
A review and catalogue of INTEGRAL-detected high-mass X-ray binaries (HMXBs) is presented by \citet[][]{2019NewAR..8601546K}. HMXBs are typically neutron stars (although black hole systems also exist), accreting material from a ${\gtrsim}5$M$_{\odot}$ companion. They are among the brightest X-ray sources in the Galaxy, as such the sample should have high completeness across the region being considered. As for the LMXBs, where available, distances are estimated from {\it Gaia DR2} parallaxes. The projected locations therefore reflect both actual HMXB locations and survey/catalogue biases (the impact of which will be quantified in Section \ref{sec:flight}). When a distance is uncertain and a range is given, we use the mean of the upper and lower estimates. As for the LMXBs, we do not separate black hole and neutron star systems.

\begin{figure*}
	\includegraphics[width=\textwidth]{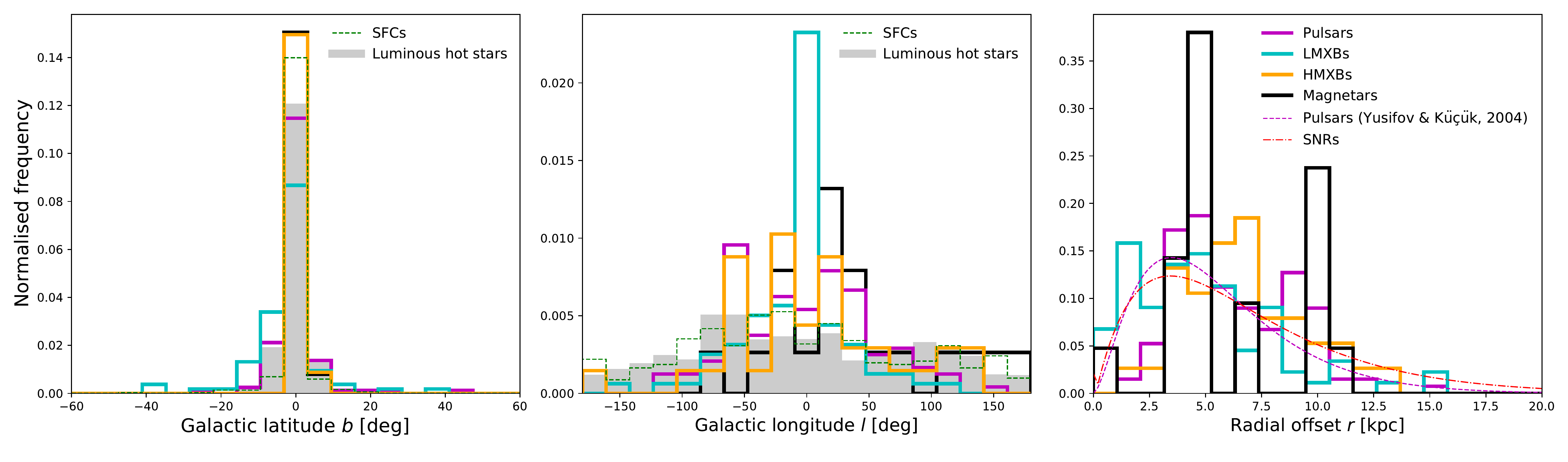}
	\vspace{-0.5cm}
    \caption{Normalised distributions in Galactic latitude, longitude and radial offset for the four Galactic neutron star samples described in Section \ref{sec:mappingNS}, with cuts applied. These are compared to star forming complexes \citep[SFCs,][]{2003A&A...397..133R}, supernova remnants \citep[SNRs][]{2021arXiv210316973V}, hot luminous stars \citep{2021A&A...650A.112Z} and the pulsar distribution of \citet{2004A&A...422..545Y}. The distributions of the neutron star samples are generally in good agreement with the previous works shown here, and others as discussed in the text. Perhaps the most prominent difference is the excess - particularly of magnetars - at 9-10\,kpc, above the smooth \citet{2004A&A...422..545Y} and \citet{2021arXiv210316973V} profiles, possibly associated with the Perseus arm.}
    \label{fig:overview}
\end{figure*}

\subsection{Comparison to other Galactic population studies}
We now discuss the broad spatial distributions of our final samples in the context of previous mapping efforts. Distributions in Galactic latitude $b$, longitude $l$ and radial offset $r$ from Galactic centre are shown in Fig. \ref{fig:overview}. Also shown on the $b$ and $l$ panels are star forming complexes from \citet{2003A&A...397..133R}, and the hot luminous star catalogue of \citet{2021A&A...650A.112Z}, which is restricted to the local 3-4\,kpc. \citet{2006ApJ...643..332F} and \citet{2004A&A...422..545Y} studied the Galactic pulsars population and found similar $b$ and $l$ distributions to both the pulsar sample shown here and young stars/SFCs, as expected.

The LMXB study of \citet{2004MNRAS.354..355J} produces a similar range in $b$ (with the majority in the central 15 degrees) and $l$ (concentrated in the 30 degrees towards GC). \citet{2021arXiv210202615A} show that LMXBs do not correlate with spiral arm structure, this can be seen in Fig. \ref{fig:overview} in their smooth $l$ distribution compared to magnetars, pulsars and HMXBs. A number of studies have noted HMXB correlations with spiral arms and SFCs \citep{2007A&A...467..585B,2012ApJ...744..108B,2013ApJ...764..185C,2021arXiv210202615A}, which can be seen here in $l$ distributions of the HMXB (and magnetar and pulsar) samples, where over-densities correspond to tangents to the spiral arms. These samples also have slightly increased numbers towards central longitudes compared to SFCs and hot stars, which may be an artefact of X-ray and radio surveys spending more time in this region.

On the right hand panel of Fig. \ref{fig:overview}, we show the Galactocentric radial offset distribution for the four neutron star samples, plus the pulsar distribution of \citet{2004A&A...422..545Y} and the supernova remnant distribution inferred by \citet{2021arXiv210316973V}. Our \citet{2005AJ....129.1993M} derived pulsar sample appears shifted to higher offsets by ${\sim}$1\,kpc compared to \citet{2004A&A...422..545Y}, similar to the simulated distribution of \citet{2006ApJ...643..332F}. The SNR distribution, which should in principle trace the locations of young neutron stars, is similarly more centrally concentrated. However, for the magnetars ($N=20$) and HMXBs ($N=36$) in particular, the samples are small enough that Poisson uncertainties may be producing noticeable differences between these distributions. The largest discrepancy is the excess of magnetars at 9-10\,kpc, over the smooth \citet{2004A&A...422..545Y} and \citet{2021arXiv210316973V} curves. While this may partly be due to low number statistics, an over-density here could also be due to the Perseus arm. Nevertheless, the final half-Galaxy samples summarised in Table \ref{tab:samples} broadly follow the expected trends when compared to previous studies of various Galactic populations.

\section{Building a Face-on Milky Way Image}\label{sec:mappingMW}
In this section we construct an image of the Milky Way in two photometric bands, on which the four neutron star populations can be placed, and measured in the same way as extragalactic transients on their hosts. As in the previous section, we restrict the mapping to $y<8.3$\,kpc. This is to match the spatial extent of the neutron star samples, and because the extrapolation of the local spiral arm structure to the far side of the disc is highly uncertain. We now describe the construction of the half-Galaxy image from three core components - the spiral arms, a smooth underlying disc, and the bulge/bar.

\subsection{Spiral arms}\label{sec:arms}
In order to map out where ongoing star formation is occurring in the Milky Way, we follow the approach of \citet{2019ApJ...885..131R}, who use water and methanol masers to trace out the spiral arm structure in 3D. Masers are uniquely suited to this, as they are numerous, bright, and their detection is relatively unaffected by foreground absorption. In brief, \citet{2016ApJ...823...77R} and \citet{2019ApJ...885..131R} use parallax measurements of various maser sources \citep[star forming complexes, young stellar objects, AGB stars and others,][]{2001A&A...368..845V,2005A&A...432..737P,2017MNRAS.469.1383G,2012ApJ...754...62A,2014MNRAS.437.1791U} to infer the spiral arm positions and parameters locally, including pitch angle and tightness. The arms are then extrapolated around the entire Galaxy. Because radio observations of masers yield radial velocities, which can be converted into a local standard of rest velocity $V_\mathrm{LSR}$, the velocity structure of the Galaxy can be mapped out to distances beyond where parallaxes can be measured. They develop a Bayesian code, with the estimated arm positions as a prior, that can infer the likeliest distance given only a Galactic latitude, longitude and LSR velocity. The warp of the disc is accounted for in the model \citep[see][]{2016ApJ...823...77R}. The code includes a prior (0 to 1) for whether the source is near or far, as the V$_{\mathrm{LSR}}$ measurements contain a distance degeneracy for sightlines along a circular path. 

We follow the same methodology as \citet{2019ApJ...885..131R} but restrict ourselves to the Red MSX catalogue of \citet[][see also \citet{2013ApJS..208...11L}]{2014MNRAS.437.1791U}. Where a near or far distance is favoured in the \citet{2014MNRAS.437.1791U} catalogue, we use P$_\mathrm{far} = $0 or 1 as appropriate, otherwise P$_\mathrm{far} = $0.5. Although less Galactic structure is mapped out by using a single input catalogue, the luminosity distribution is well understood, as is the incompleteness as a function of distance. Limiting ourselves to a single, well understood catalogue also avoids double counting star forming complexes or single sources that appear in multiple maser catalogues. We further limit the sample to HII regions and young stellar objects (YSOs), in order to trace out star formation specifically. The resultant star forming region catalogue comprises 1644 sources.

Taking the 1644 HII region and YSO masers of \citet{2014MNRAS.437.1791U}, we run the Bayesian distance {\sc fortran} code of \citet{2019ApJ...885..131R} and obtain a Milky Way spiral arm map. Beyond 3\,kpc, the completeness of the input catalogues drops off as a function of distance from the Solar System (inverse square law) and Galactocentric radius (mainly due to source confusion towards the Galactic centre). We want to correct for this, so that the the following three requirements are satisfied:
\begin{itemize}
    \item The mean distance from each HII region/YSO maser to the next nearest neighbouring maser as a function of Galactocentric radius does not depend on Galactocentric azimuthal angle,
    \item the mean number of masers within a fixed distance of each maser, again as a function of Galactocentric radius, does not depend on Galactocentric azimuthal angle,
    \item the luminosity function of the masers is the same across the map.
\end{itemize}
To compensate for the high level of incompleteness at larger distances, we aim to match the density of sources in the local (${\sim}$3\,kpc) neighbourhood according to the above metrics, across the (half) Galaxy.

We again follow the method of \citet{2019ApJ...885..131R}. Artificial masers (sprinkles) are added at random $x$ and $y$ offsets from each of the 1644 {\it real} maser positions. The number of sprinkles added per real HII region/YSO maser varies as a function of distance from Earth and the Galactocentric radius. More random sprinkles are added at larger distances, and the width of the Gaussian used to determine the random $x$ and $y$ offsetss from the existing maser positions increases with Galactocentric radius (to reflect the widening spiral arm width). This process is described by the following. The number of sprinkles added for each of the 1644 masers is given by,
\begin{equation}
    N_\mathrm{sprinkle}=\Big(\frac{D}{3\mathrm{kpc}}\Big)^{2} - N_\mathrm{corr}
	\label{eq:sprinkle}
\end{equation}
at distances greater than 3\,kpc. Below this, no sprinkles are added. The maximum number added is 10, and the minimum 1. $N_\mathrm{corr}$ is a correction to the form given by \citet{2019ApJ...885..131R}, which is necessary given our smaller input catalogue size. The $N_\mathrm{sprinkle}$ sources are added following Gaussians in $x$ and $y$. The standard deviation in each is given by,

\begin{equation}
    \sigma_{xy}= \frac{336 + 36(R_{g}-8.15)}{W_\mathrm{corr}}
	\label{eq:xygaussian}
\end{equation}
where $W_\mathrm{corr}$ is a correction we have added, again needed in order to tune the random sprinkling to match the local 3\,kpc density of points. In \citet{2019ApJ...885..131R}, with their large input catalogue, $N_\mathrm{corr} = 0$ and $W_\mathrm{corr} = 1$.

A final issue is that there is a dearth of sources in a cone of around ${\sim}$30 degrees toward the Galactic centre. On our side of the Galactic centre, this is corrected for by fitting a parabola to join the gap in the arms and sprinkling points along these curves so that their spatial density is comparable to the arms either side of the gap. No correction is made to the far side of the disc, as we only consider the half-Galaxy at $y<8.3$\,kpc going forwards.

\begin{figure}
	\includegraphics[width=\columnwidth]{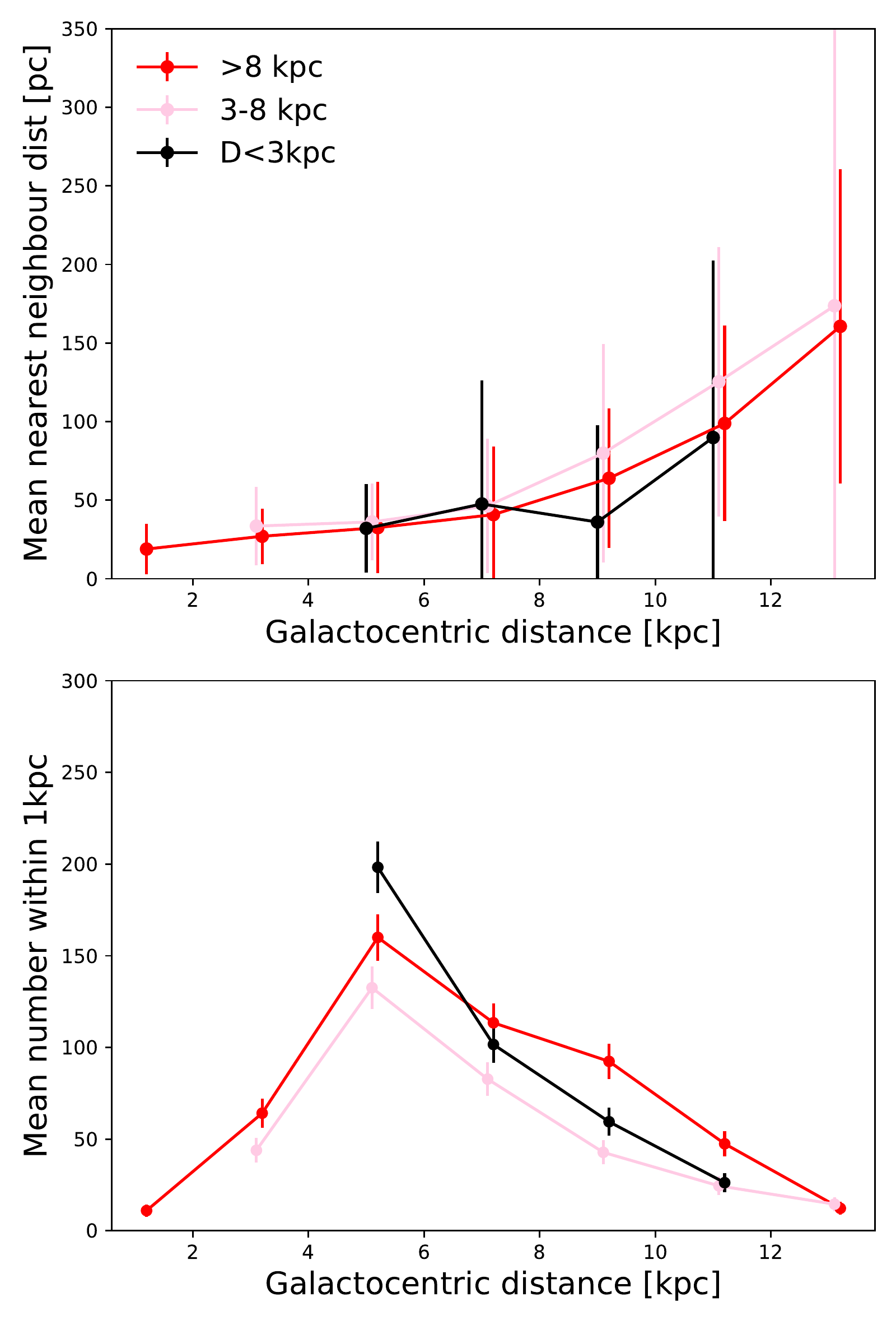}
    \caption{Upper panel: the mean distance to the nearest neighbouring maser, following the method of \citet{2019ApJ...885..131R} as described in the text. The 3 heliocentric distance ranges have approximately equal maser density as a function of Galactocentric distance. Lower panel: The mean number of masers within 1\,kpc of each maser, again as a function of radial distance from Galactic centre. The arm width and N$_\mathrm{sprinkle}$ functions used to add artificial masers are given in equations \ref{eq:sprinkle}, and \ref{eq:xygaussian}.}
    \label{fig:sprinklefitting}
\end{figure}

\begin{figure*}
	\includegraphics[width=\textwidth]{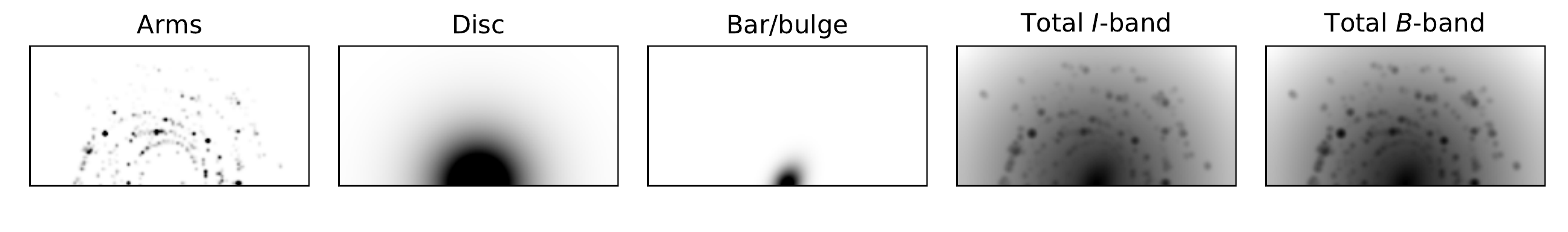}
	\vspace{-0.5cm}
    \caption{The three Galaxy components used, and their combination (with appropriate scaling) to produce the $I$ and $B$-band images. The spiral arm locations are derived from YSO and HII region masers in the \citet{2014MNRAS.437.1791U} catalogue, assigned a distance using the \citet{2019ApJ...885..131R} distance calculator. The disc is an exponential with scale length 2.6\,kpc, and the bulge is a S\'ersic profile convolved with the bar profile of \citet{2020MNRAS.492.3128G}, at an angle of 27 degrees to the Sun-Galactic centre line.}
    \label{fig:components}
\end{figure*}

In order to tune the incompleteness corrections described by equations \ref{eq:sprinkle} and \ref{eq:xygaussian}, so that the first two bullet-point criteria above are met, we vary $N_\mathrm{corr}$ and $W_\mathrm{corr}$. The Galaxy is split into three regions: sources within 3\,kpc, those in the helicoentric distance range $3-8$\,kpc, and those beyond 8\,kpc. In Fig. \ref{fig:sprinklefitting}, the target radial profiles are shown in black, representing the maser distribution in the local 3\,kpc. We aim to find $N_\mathrm{corr}$ and $W_\mathrm{corr}$ values such that, when the artificial maser sprinkles are added, the $3-8$\,kpc and $>8$\,kpc profiles match those at $<3$\,kpc. This will correct for incompleteness, so that it is at least uniform across the half-Galaxy region in consideration. We find that the default $N_\mathrm{corr} = 0$ and $W_\mathrm{corr} = 1$ values from \citet{2019ApJ...885..131R} provide a poor agreement between local sources and the rest of the Galaxy, unsurprising given our smaller input catalogue. Sampling values in the range 0-8, $N_\mathrm{corr} = 6$ and $W_\mathrm{corr} = 5$ provide the best match in terms of ${\chi}^{2}$ between the local and more distant regions. The result is shown in Fig. \ref{fig:sprinklefitting}.

Each sprinkle is also assigned a bolometric luminosity. Inside 3\,kpc, the \citet{2014MNRAS.437.1791U} catalogue is complete down to bolometric luminosities of 1000L$_{\odot}$. Therefore, outside 3\,kpc, the randomly sprinkled points are assigned a luminosity drawn randomly from the distribution below the completeness limit at that distance. In this way, the luminosity function is approximately constant across the half-Galaxy map, complete down to 1000L$_{\odot}$ everywhere. 

The resultant maser map is shown in the left-hand panel of Fig. \ref{fig:components} (and can also been seen in the final image, see Fig. \ref{fig:milkymap}). To compare the arm positions to other Milky Way spiral arm maps, references include \citet{2002ApJ...566..261V}, \citet{2003A&A...397..133R}, \citet{2006ApJ...643..332F}, \citet{2008AJ....135.1301V}, \citet{2009PASP..121..213C}, in addition to more local mapping efforts using {\it Gaia} \citep[e.g.,][]{2021arXiv210504590C,2021A&A...650A.112Z} and electron density models \citep[e.g.][]{2002astro.ph..7156C}.

\subsection{The disc, bulge and bar}
In addition to the arms, the Milky Way has a disc (in which the arms are embedded) and a central bulge/bar. Our basic model follows \citet{2008ApJ...679.1239W} and \citet{2016ARA&A..54..529B}. We model the disc with an exponential profile, adopting a scale length of R$_{d} = 2.6$\,kpc. This value is based on the range in the literature, and is around the mean of thin disc (which dominates the disc stellar content) values. The bulge is modelled as a S\'ersic profile with index $n=1.32$ and half-light radius R$_{d} = 0.64$\,kpc \citep{2008ApJ...679.1239W}. To account for the bar, this is convolved with the bar profile as mapped by Mira variables in {\it Gaia} \citep{2020MNRAS.492.3128G}. The angle between the Sun-Galactic centre line and bar semi-major axis is taken to be 27 degrees \citep[e.g.][]{1997MNRAS.288..365B,2016ARA&A..54..529B}.

\subsection{Weighting the components}
The three components - a smooth disc, embedded arms and bulge/bar - now need weighting to represent their respective contributions to the total luminosity of the Milky Way. This will also be wavelength dependent.

As a starting point, we take the bulge and disc luminosities of \citet{2006MNRAS.372.1149F}, where the disc includes the arms. The $I$-band bulge and disc luminosities are 10$^{10}$L$_{\odot}$ and 3$\times$10$^{10}$\,L$_{\odot}$ respectively. 

The procedure described in Section \ref{sec:arms} means that we have a value for the total bolometric luminosity of the masers above the 1000L$_{\odot}$ limit, 1.78$\times$10$^{8}$L$_{\odot}$. This is twice the \citet{2014MNRAS.437.1791U} estimate for the total contribution of high mass embedded star formation to the Galactic luminosity (due to differences in incompleteness corrections). However, for our purposes we are only interested in the spiral arm locations, and the {\it relative} flux contributions of the galaxy components.

To determine what fraction of the disc light should arise from the spiral arms versus the smooth underlying exponential, we adopt an arm strength of 0.15 in the $I$-band, typical for Milky Way mass galaxies \citep[e.g.][]{2019A&A...631A..94D,2020ApJ...900..150Y}. Arm strength is defined as the ratio of disc surface density arising from the $m=2$ spatial Fourier component, versus the $m=0$ component \citep{1998MNRAS.299..685S}. It therefore quantifies the proportion of disc light arising from short spatial scales (i.e., the spiral arm structure). Arm strength varies in different bands, as the spiral arms are usually slightly bluer than the inter-arm regions - for $I$-band strengths of 0.15, typical $B$-band strengths are ${\sim}$0.2, values which we use going forward \citep{2018ApJ...862...13Y}. 

We now assume that $I$-band luminosity of the maser-associated HII regions is proportional to their maser-inferred bolometric luminosity. If the $I$-band arm strength is 0.15, the $I$-band exponential disc must have a luminosity $\sim5.5$ times that of the arms. The $I$-band bulge/bar luminosity is then one third of the disc$+$arm total (as the \citet{2006MNRAS.372.1149F} disc value includes spiral arms). The constants in the exponential disc and  S\'ersic profiles are scaled to reflect these respective contributions to the $I$-band light.

We now have an $I$-band image with the components appropriately scaled. However, the $I$-band is a better tracer of stellar mass than star formation. To obtain a bluer version of the image, we use the $B$--$I$ colours of \citet{2015ApJ...809...96L}, who estimate a total integrated Galaxy colour - including dust - of 1.77. \citet{2016ARA&A..54..529B} obtain an alternative value of 1.85$\pm$0.1, but this is consistent with \citet{2015ApJ...809...96L}. Note that all $B$ and $I$ magnitudes and colours quoted in this paper are in the Vega system, but they are converted to AB magnitudes to perform any calculations. 

We now need the $B$--$I$ colour of either the disc or bulge in order to determine the $B$-band luminosities of all three components. \citet{2006MNRAS.372.1149F} provide a local disc $B$--$I$ colour of 1.48, but this has been corrected for dust and is therefore artificially blue. We instead use the `as observed' (not dust-corrected) Milky Way colour estimates of \citet{2015ApJ...809...96L}, which are based extragalactic analogues chosen for their mass and SFR similarity with the Milky Way. Within these analogues they identify a bulge-dominated sub-sample, which we use to estimate the Milky Way bulge colour. This has a mean $B$-band mass-to-light ratio (MLR) of 4.1. We also use their global MLR estimates in $I$ and $B$ of 1.29 and 1.89, and a bulge-to-total mass ratio of 0.3 \citep{2016ARA&A..54..529B}. Taking the $I$-band component luminosities from \citet{2006MNRAS.372.1149F}, this implies a bulge/bar $B$--$I$ of 2.41. 

We now have a bulge/bar $B$--$I$ of 2.41 a global $B$--$I$ of 1.77, and $I$-band luminosities for the total Galaxy and the components. We can therefore infer the colour of the final component (disc including arms) - 1.62 - and hence we have the luminosities of all three components in each band. This disc colour is bluer than the total integrated colour as expected, and 0.14 redder than the \citet{2006MNRAS.372.1149F} dust-corrected value for the local disc.

\subsection{The final Milky Way image}
The construction of the final Galaxy image is shown in Fig. \ref{fig:components}, in both bands. These images have a pixel scale of 250\,pc/pix, and we have applied a Gaussian blur with a FWHM of 2 pixels to simulate {\it Hubble Space Telescope, (HST)} quality imaging (${\sim}$0.05\,arcsec\,pix$^{-1}$). The masers are treated as point sources, and their luminosity is assigned to whichever pixel they fall under. In reality they are not point sources, but this unlikely to be a significant issue, as the majority of star forming regions have radii $<250$\,pc, and almost all $<500$\,pc \citep{2012MNRAS.422.3339W}. This effect is further diminished as they tend to have approximately Gaussian density profiles, so that the light contributed to neighbouring pixels will not usually be a significant fraction of the total star forming region luminosity. The final $B$ and $I$-band images are similar, as the three input components have the same profiles, but the different scaling of these components with respect to each other means that there are differences (e.g. the light is more centrally concentrated in the $I$-band image). The final half-Galaxy $B$ and $I$-band images are available in fits format at \url{https://github.com/achrimes2/MW-NS-Flight}.

\subsection{Placing Galactic neutron stars on the image}
\begin{figure*}
    \centering
    \includegraphics[width=0.99\textwidth]{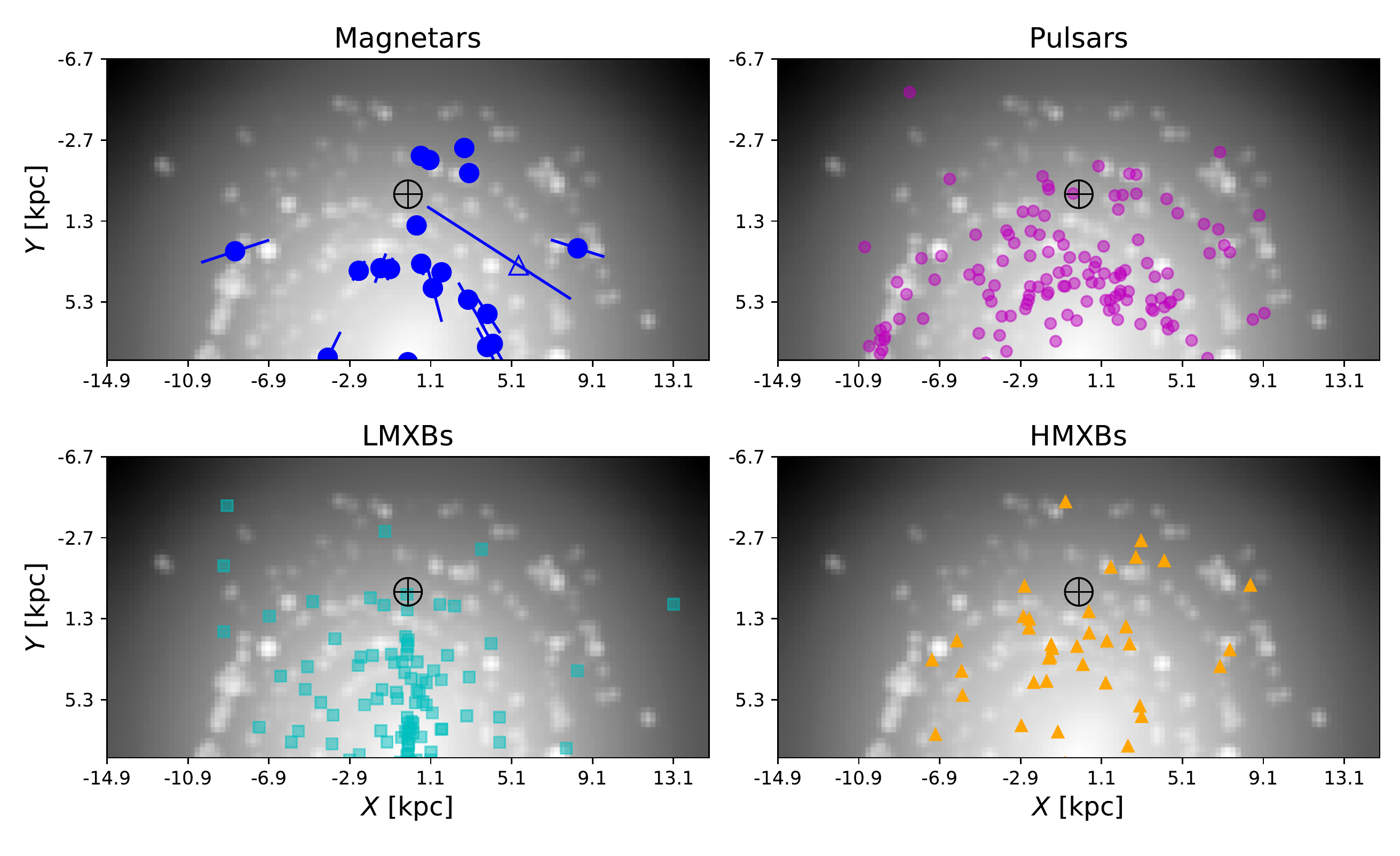}
    \caption{The Milky Way map overlaid with magnetars positions \citep[top left, blue dots with distance uncertainties indicated,][]{2014ApJS..212....6O}, pulsars \citep[top right, magenta dots,][]{2005AJ....129.1993M}, LMXBs \citep[bottom left, cyan squares,]{2020NewAR..8801536S} and HMXBs \citep[bottom right, orange triangles,][]{2019NewAR..8601546K}. The Solar system is located at (0,0), and Galactic centre at (0,8.3). The FRB source SGR\,1935+2154 is indicated in the magnetar panel by a triangle. Only the half-Galaxy region shown is used for measurements throughout, to limit the impact of heliocentric distance uncertainties and detection biases.}
    \label{fig:milkymap}
\end{figure*}

The final neutron star samples compiled in Section \ref{sec:mappingNS} are now placed on the half-Galaxy image constructed in Fig. \ref{fig:components}. Their positions have been computed from the Galactic latitude, longitude, and given distance, correcting for a Solar System height above the Galactic plane of 15\,pc \citep[][ although this negligible for our purposes]{2014ApJS..212....6O}. Their positions on the $B$-band image are shown in Fig. \ref{fig:milkymap}. The selected magnetars, bright pulsars, INTEGRAL LMXBs and HMXBs are shown as blue circles, magenta circles, cyan squares and orange triangles respectively.

\section{Offsets and Enclosed fluxes}\label{sec:offsets}
\subsection{Offsets}
Having selected our neutron star samples in Section \ref{sec:mappingNS}, and constructed Milky Way model in Section \ref{sec:mappingMW}, we now move on to measuring how neutron stars are distributed with respect to Galactic structures and light. We start with their offsets from Galactic centre.

\begin{figure*}
	\includegraphics[width=0.99\textwidth]{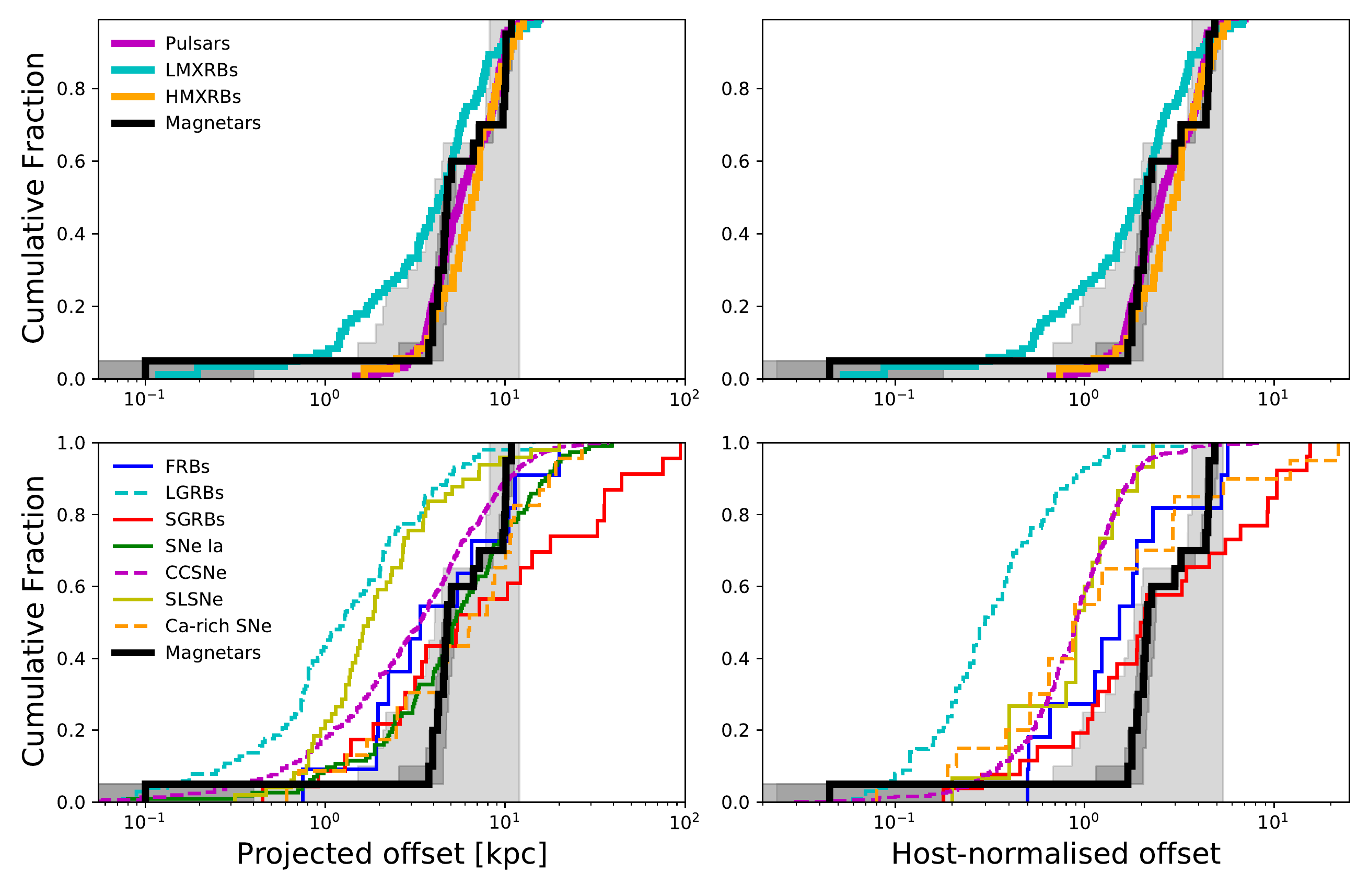}
    \caption{Projected (left) and host normalised (right) offsets for Milky Way magnetars, compared to other Galactic neutron star systems (top row) and the offsets of extragalactic transients on their hosts (bottom row, shown separately for clarity). FRB\,20200120E has been added to the FRB offset dataset, assuming association with M81 \citep{2021arXiv210301295B}. The light grey shaded region is bounded by the extreme upper and lower distributions for the magnetars, assuming the maximum and minimum offsets possible along that line of sight (and that the magnetar lies `in' the Galaxy, at ${\delta}r<12$\,kpc). The narrower dark grey shading indicates the maximum and minimum allowed within the quoted magnetar distance uncertainties. A 15 per cent error is used where one is not available. The $I$-band host-normalised offset is shown (with half-light radius r$_{h} = 2.22$\,kpc), similar results are obtained using the $B$-band r$_{h}$ of 2.37\,kpc (see Table \ref{tab:info} of appendix \ref{apx:A}). }
    \label{fig:offsets}
\end{figure*}

We now look at the offsets of the neutron stars from the Galactic centre. Our offsets measurements are 2D projections of the 3D distance from Galactic centre. Both offsets ${\delta}r$ and host-normalised offsets ${\delta}r/r_{e}$ are shown in the upper panels of Fig. \ref{fig:offsets}, where $r_{e}$ is the half-light radius. For our Milky Way image, $r_{e} = 2.37$\,kpc in the $B$-band and $2.22$\,kpc in the $I$-band. In principle, normalising by the half-light radius should make comparisons between galaxies of difference physical sizes fairer.
 
The four Galactic samples are compared to extragalactic transient offsets distributions, shown in the lower two panels of Fig. \ref{fig:offsets}. The comparison data are from \citet[][FRBs]{2020arXiv201211617M}, \citet[][LGRBs]{2016ApJ...817..144B,2017MNRAS.467.1795L}, \citet[][SGRBs]{2013ApJ...769...56F}, \citet[][type Ia SNe]{2020ApJ...901..143U}, \citet[][CCSNe]{2012ApJ...759..107K,2020arXiv200805988S}, \citet[][SLSNe]{2015ApJ...804...90L,2020arXiv200805988S}, and \citet[][Ca-rich SNe]{2017ApJ...836...60L,2020ApJ...905...58D}. We have also added the M81 repeater to the FRB distribution \citet{2021arXiv210301295B}. The results of AD-tests between the magnetars/pulsars/XRBs offsets and the comparison samples are listed in Table \ref{tab:ADoffset}. Results which do not round up to 0.01 are simply listed as 0.00. The maximum value is 0.25, as outputs are capped at this value by the {\sc scipy} Anderson function. 

Across the four Galactic samples, FRBs most frequently have a p-value $>0.05$. Only the $I$-band normalised results are shown in Fig. \ref{fig:offsets}, using the $B$-band half-light radius results in slightly less offset distributions (we refer the reader to appendix \ref{apx:A}, where the full magnetar results are listed in Table \ref{tab:info}, and access to the data for the other samples is described).

To quantify the uncertainty on the magnetar distribution, we take the the maximum and minimum offsets possible along the line of sight, assuming that the magnetars lie `inside' the Galaxy (i.e. $<$12\,kpc from Galactic centre). The region bounded by the minimum and maximum possible distributions is shaded light grey in Fig. \ref{fig:offsets}. We also take the listed uncertainties, shading between the quoted upper and lower bounds produces the narrower dark grey band. Even if we consider the lowest offsets that the magnetars could possibly have along their sightlines, they cannot match the CCSN offset distribution.

\begin{table}
\centering 
\caption{Offset and host-normalised offset AD test results, for the comparisons made in Fig. \ref{fig:offsets}. Top: offsets, middle: $B$-band host normalised offset (the comparison data is also normalised in a UV/blue band where available), bottom: $I$-band host normalised offset (comparison data normalised in an $r$ or $I$-band). p-values of 0.05 or greater, indicating statistical consistency, are highlighted in bold and have a grey background. None of the Galactic populations are consistent with LGRBs, SLSNe or CCSNe according to this measure. Note that host-normalised data for SNe Ia are not available.} 
\label{tab:ADoffset}
\begin{tabular}{lllllll}
\hline %
          & LGRB & SLSN & CCSNe & FRB                                   & SNe Ia                                 & SGRB    \\
\hline %
Offset \\
\hline %
Magnetars & 0.00 & 0.00 & 0.00  & \cellcolor{shade}\textbf{0.07} & \cellcolor{shade}\textbf{0.17} & 0.00                                  \\
HMXRBs    & 0.00 & 0.00 & 0.00  & 0.02                                  & 0.02                                  & 0.00                                  \\
Pulsars   & 0.00 & 0.00 & 0.00  & 0.00                                  & 0.00                                  & 0.00                                  \\
LMXRBs    & 0.00 & 0.00 & 0.04  & \cellcolor{shade}\textbf{0.25} & 0.01                                  & 0.00                                  \\
\hline 
$B$-band normed \\
\hline 
Magnetars & 0.00 & 0.00 & 0.00  & 0.04                                  & n/a                                   & \cellcolor{shade}\textbf{0.08} \\
HMXRBs    & 0.00 & 0.00 & 0.00  & 0.00                                  & n/a                                   & 0.01                                  \\
Pulsars   & 0.00 & 0.00 & 0.00  & 0.00                                  & n/a                                   & 0.00                                  \\
LMXRBs    & 0.00 & 0.00 & 0.00  & \cellcolor{shade}\textbf{0.25} & n/a                                   & 0.02                                  \\
\hline 
$I$-band normed \\
\hline 
Magnetars & 0.00 & 0.00 & 0.00  & 0.02                                  & n/a                                   & \cellcolor{shade}\textbf{0.06} \\
HMXRBs    & 0.00 & 0.00 & 0.00  & 0.00                                  & n/a                                   & 0.01                                  \\
Pulsars   & 0.00 & 0.00 & 0.00  & 0.00                                  & n/a                                   & 0.00                                  \\
LMXRBs    & 0.00 & 0.00 & 0.00  & \cellcolor{shade}\textbf{0.25} & n/a                                   & 0.03                 \\
\hline %
\end{tabular}
\end{table}

Looking at Fig. \ref{fig:offsets}, it is evident that the specific morphology of the Milky Way dictates the shape of the Galactic distributions. The choice of disc scale length has an impact of the normalised offsets, but literature estimates only vary by at most a factor of ${\sim}$1.5-2 \citep{2016ARA&A..54..529B}. Much of the light is concentrated centrally in the bulge (in both bands), a predominantly older population where fewer young neutron stars reside. Consequently, the LMXBs (known to be a older population) appear distributed on the Milky Way's $I$-band light in a similar way to supernovae on their hosts, although this does not imply a connection. Likewise, the younger pulsars, HMXBs and magnetars are more offset than the short-delay time extragalatic transients. This arises because we are dealing with a single Galaxy: in the extragalactic samples, the range of host morphologies, sizes and viewing angles removes any such trends.

There may also be issues related to faint, halo light that is not detected in extragalactic samples \citet{2021arXiv210111622P}, and other Galaxy-specific morphological effects that influence the offsets and flux profile, such as the specific arm locations. For example, the Galactic ring of star formation at 3-5\,kpc has a high density of magnetars, relative to other arm structures. This is reflected as a dearth of objects at $<3$\,kpc and a rapid climb in the cumulative distribution at $3-5$\,kpc. Interestingly, magnetar locations appear to favour this inner, nuclear ring, over spiral arms further out. This does not appear to be a distance-sensitivity effect as similarly bright arms at comparable heliocentric distances still have fewer magnetars. This may be suggestive of other factors playing a role in their production, beyond a high star formation rate (such as metallicity or IMF variations) and warrants further investigation.

\subsection{Enclosed fluxes}
\begin{figure}
	\includegraphics[width=\columnwidth]{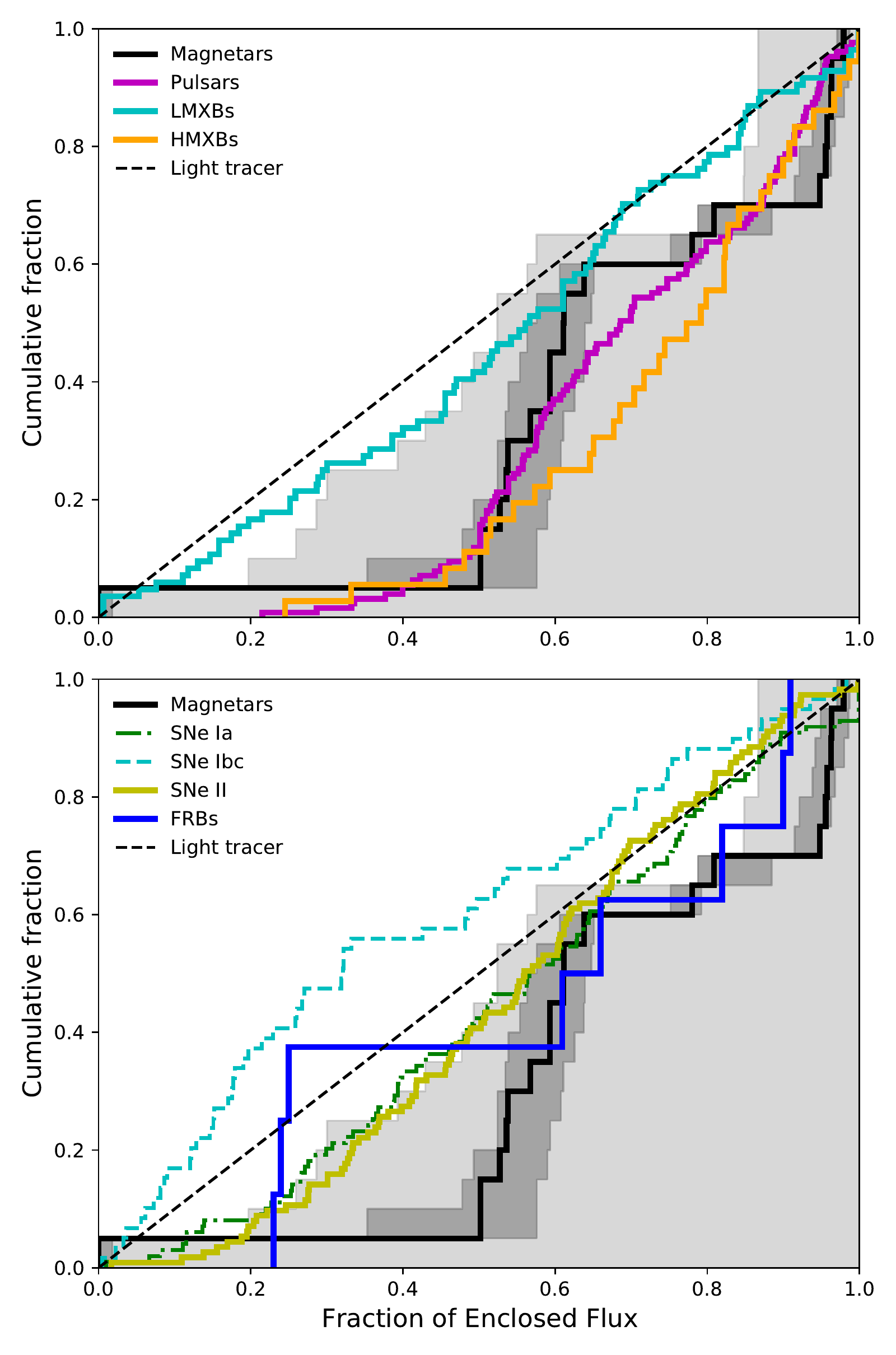}
    \caption{Top: enclosed fluxes for the four Galactic samples. Bottom: the fraction of Milky Way flux (at $y<8.3$\,kpc) enclosed within the Galactocentric radius of the magnetars, the $I$-band version is used for a fairer comparison to the extragalactic samples. The maximum and minimum distributions possible for the magnetars are shaded in light grey, the same is shown in dark grey using their given distance uncertainties (or a 15 per cent error where one is not available). }
    \label{fig:encflux}
\end{figure}

Another way to measure how the neutron stars relate to light is to measure the fraction of the total Galactic flux enclosed at their radial distance. We calculate this for the magnetar, pulsar and XRB samples, comparing to supernova \citep{2009MNRAS.399..559A,2015PASA...32...19A} and FRB distributions \citep{2020arXiv201211617M} in Fig. \ref{fig:encflux}. The grey shaded regions are the magnetar uncertainties, calculated in the same way as for the offsets. We again restrict the samples and pixels to $y<8.3$\,kpc, and measure the fraction of flux enclosed in the semicircle produced.

\begin{table}
\centering 
\caption{As for Table \ref{tab:ADoffset}, but for enclosed fluxes. The distributions are shown in Fig. \ref{fig:encflux}. Only the $I$-band is used for a fair comparison to the extragalactic samples.} 
\label{tab:ADencflux}
\begin{tabular}{llllll}
\hline %
           & Type Ia SNe                        & FRB                                & SNe II                        & SNe Ibc                       \\
\hline %
Magnetars & 0.04                                  & \cellcolor[HTML]{DDDDDD}\textbf{0.25} & 0.02                                  & 0.00         \\
HMXRBs    & 0.00                                  & \cellcolor[HTML]{DDDDDD}\textbf{0.11} & 0.00                                  & 0.00         \\
Pulsars   & 0.00                                  & \cellcolor[HTML]{DDDDDD}\textbf{0.07} & 0.00                                  & 0.00         \\
LMXRBs    & \cellcolor[HTML]{DDDDDD}\textbf{0.25} & \cellcolor[HTML]{DDDDDD}\textbf{0.25} & \cellcolor[HTML]{DDDDDD}\textbf{0.25} & 0.01           \\
\end{tabular}
\end{table}

AD-test results are given in Table \ref{tab:ADencflux}. FRBs are the transient most often consistent with the various neutron star populations. The enclosed flux measurements, like the offsets, are biased by the specific morphology of the Milky Way, specifically because the neutron star distributions are being drawn from just one Galaxy. To disentangle these effects, we now look at the fraction of light statistic, which operates on a pixel-by-pixel basis and can be restricted to sub-regions within the Galaxy.

\begin{figure*}
  \centering
    \includegraphics[width=0.95\columnwidth]{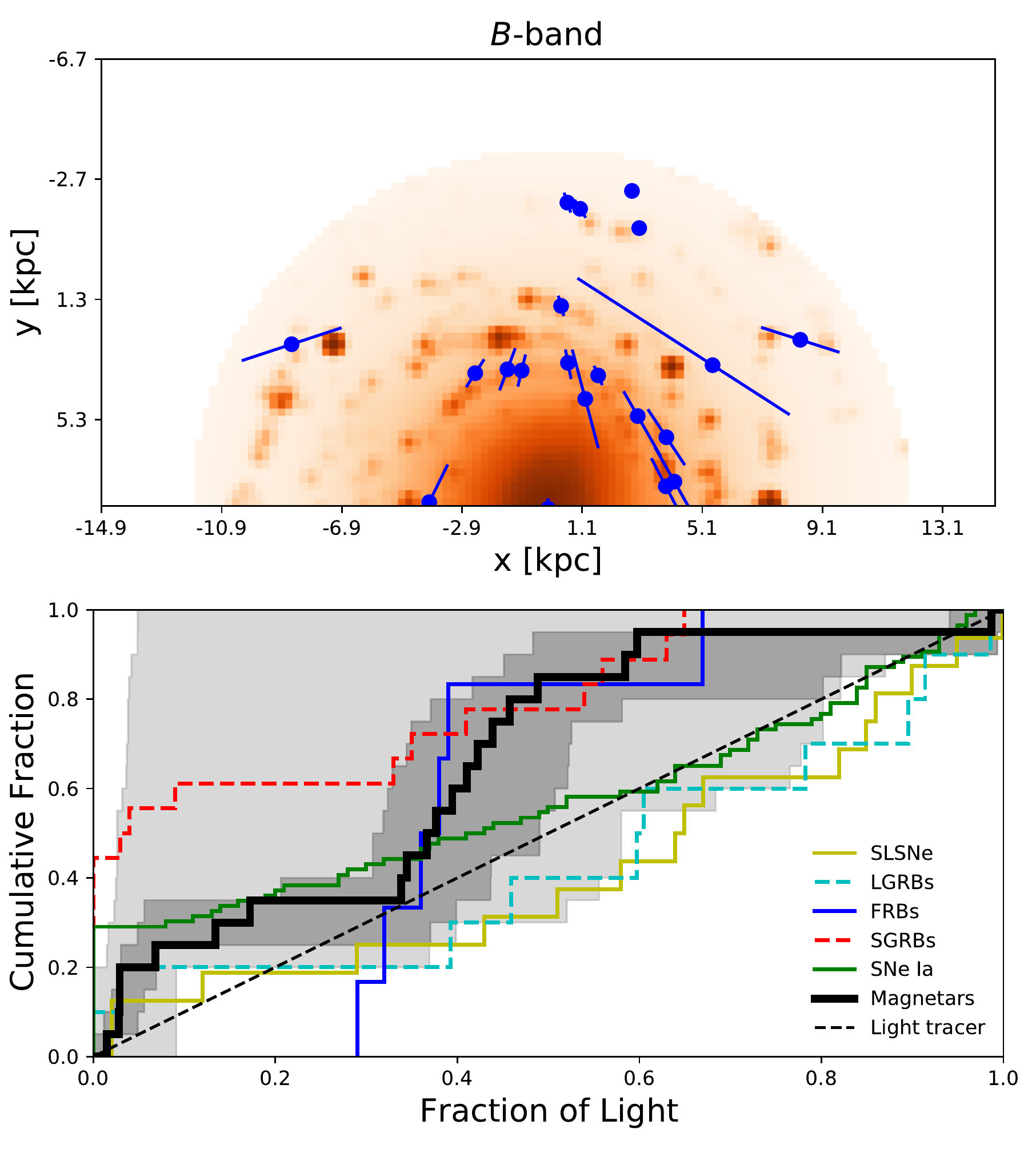}
    \includegraphics[width=0.91\columnwidth]{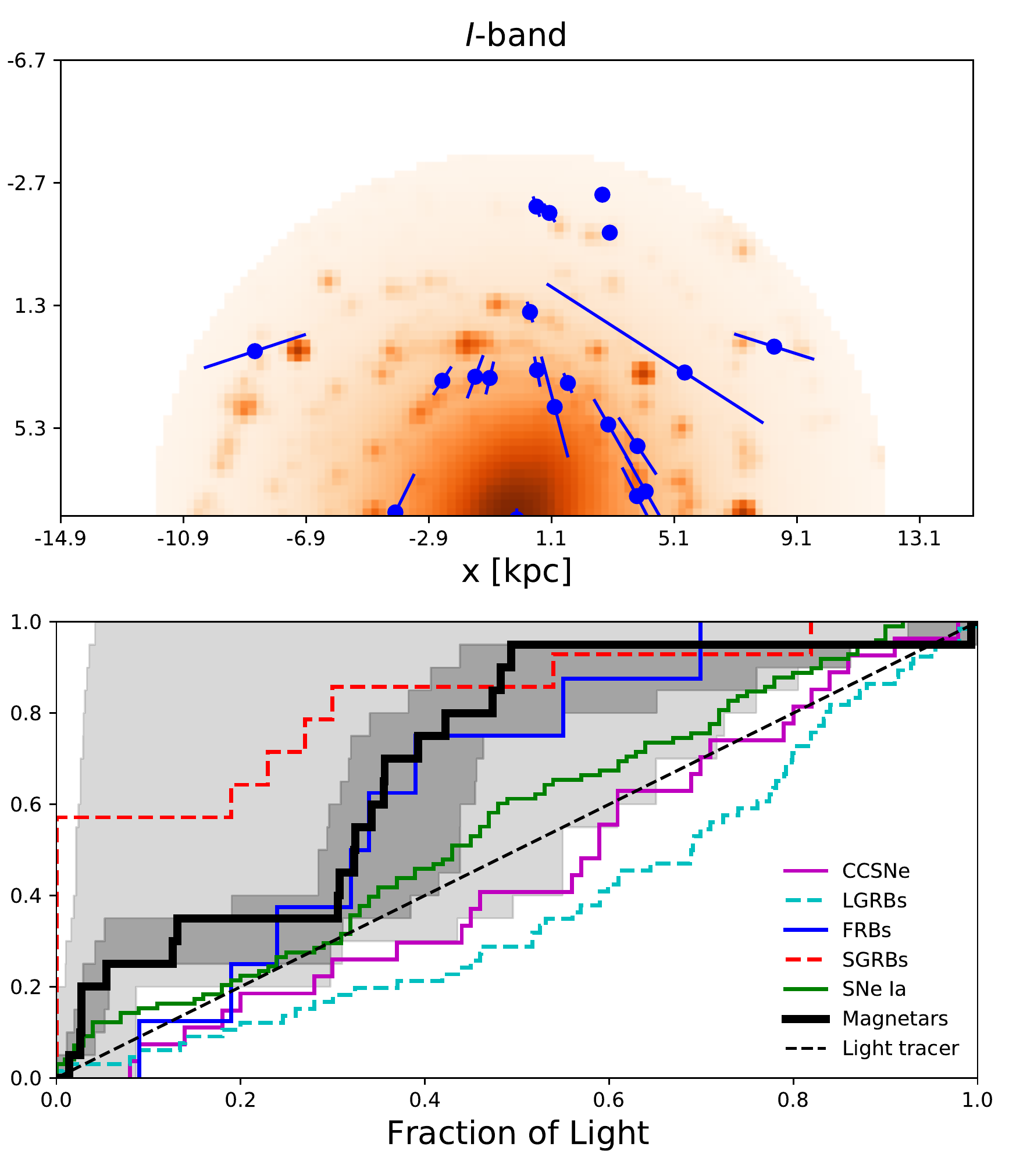}
    \caption{Upper panels: half-Galaxy F$_\mathrm{light}$ colour maps with magnetars positions and uncertainties indicated ($B$-band left, $I$-band right). The 250\,pc/pixel scale (with a 2 pixel FWHM) is broadly representative of {\it HST} spatial resolution for cosmological transients. Pixels within a 12\,kpc Galactocentric radius are selected for calculating the F$_\mathrm{light}$ distributions. Lower panels: the corresponding fraction of light distributions for the magnetars (solid black lines). The light grey shaded regions represent the minimum and maximum distributions that arise from taking the minimum and maximum F$_\mathrm{light}$ values along the magnetar sightlines. This places a strong upper limit on how concentrated magnetars can be on the Galactic light. The darker grey bands use linear sampling between the McGill catalogue upper and lower distance estimates, taking the mean uncertainty of 15 per cent on a distance where a range is not provided. The distributions for various extragalactic transients on their host galaxies are also shown.}
    \label{fig:flightBI}
\end{figure*}

\section{Fraction of light}\label{sec:flight}
We now turn to the fraction of light \citep[F$_\mathrm{light}$,][]{2006Natur.441..463F} statistic. The F$_\mathrm{light}$ distribution is calculated by ranking host-associated pixels by brightness, and assigning each their value in the cumulative sum, normalised by the total (half) galaxy cumulative flux. The brightest pixel therefore has the value 1. Transients that arise from the light in an unbiased way produce an F$_\mathrm{light}$ cumulative distribution that follows a 1:1 line, those that are concentrated on brighter regions have a distribution below the 1:1 line, and those are that offset or avoid bright regions lie above this line. UV/blue bands are assumed to trace young stars and thus star formation, whereas IR/red bands better trace stellar mass. 

The fraction of light method has the benefit of being less dependent on host morphology. For example, if a transient occurs on a bright star forming region, but at a large offset from the host centre, both the offset and enclosed flux measurements will indicate a low likelihood of association with star formation/stellar mass (depending on the wavelength). The F$_\mathrm{light}$ value, however, will assign the proportionate ranking in the distribution of host pixels, ranking it highly, and correctly identifying a close association. It is also possible to restrict the region considered, for example, if we want to look solely at the disc \citep{2017MNRAS.467.1795L}. This enables use to probe how transients (and in our case neutron stars) trace light in different stellar populations. 

Because F$_\mathrm{light}$ relies on pixel values, the choice of spatial resolution is important. The adopted resolution of 250\,pc per pixel is broadly typical of {\it HST} resolution for targets at cosmological ($z\sim1$) distances.

\subsection{Half-Galaxy results}
We select all pixels at $y<8.3$\,kpc and within 12\,kpc of the Galactic centre as being associated with the galaxy. Pixels outside this semicircular region are assigned F$_\mathrm{light} = 0$. The pixels in the half-Galaxy are ranked by their cumulative value and normalised. The pixel selection and F$_\mathrm{light}$ colour maps are shown in the upper panels of Fig. \ref{fig:flightBI}. In the lower panels, magnetar F$_\mathrm{light}$ distributions are compared with extragalactic transients. The comparison data are from \citet[][FRBs]{2020arXiv201211617M}, \citet[][LGRBs]{2016ApJ...817..144B,2017MNRAS.467.1795L}, \citet[][SGRBs]{2013ApJ...769...56F}, \citet[][type Ia SNe]{2013Sci...340..170W}, \citet[][CCSNe]{2010MNRAS.405...57S} and \citet[][SLSNe]{2015ApJ...804...90L}.

The light and dark grey uncertainty bands on the magnetar distributions are again derived from the maximum and minimum pixel values possible along the line of sight, and within the quoted distance ranges (or with a  15 per cent distance uncertainty if unavailable). Where an errorbar takes us past $y=8.3$\,kpc, or when calculating the full range of values along a sightline, we have to quantify the expected pixel values on the far side of the disc. To do this, we simply mirror the galaxy across $y=8.3$\,kpc, sampling the mirrored values out to the appropriate distance (either the upper distance estimate, or the 'edge' of the Galaxy). This should give a reasonable estimate of the range of pixel values expected for sightlines beyond $y=8.3$\,kpc, assuming the Galaxy is approximately symmetric on large scales.

The dark grey uncertainty band is somewhat larger here than for the offsets or enclosed fluxes. This is because we are now dealing with pixel values, which can have far more variation along the same line of sight than offsets (no dependence on pixel values) or enclosed fluxes (summing pixel values within a radius). We note that, because the quoted uncertainties are typically not formally quantified, we have simply shaded between the upper and lower heliocentric distances (i.e., assuming constant probability density in this range). In reality, the probability density closer to the assumed magnetar distances is likely greater than at the edges of these ranges. 

The uncertainties are slightly larger in the $B$-band (this can be seen in Fig. \ref{fig:flightBI}), because the arms contribute a higher fraction of the total flux. Random variations in location due to distance uncertainties therefore sample a wider range of pixel values along a given line of sight, providing a wider range of possible distributions. Based solely the range possible along a line-of-sight, it is possible that many magnetars actually lie on highly-ranked pixels and are not inconsistent with CCSNe or even SLSNe. However, the LGRB distribution cannot be reached, independent of the heliocentric distances assumed.

AD tests between the magnetars/pulsars/XRBs and extragalactic transients are listed in Table \ref{tab:ADflightB}. The equivalent figures for the pulsars and XRBs are available in Figs \ref{fig:helio_pulsars}, \ref{fig:helio_LMXBs} and \ref{fig:helio_HMXBs} of Appendix \ref{apx:B}. Across the four Galactic samples, as we found for the offset and enclosed flux measurements, FRBs fail to reject the null hypothesis that they are drawn from the same distribution as Galactic neutron stars the most often. SGRBs are the least consistent with Galactic neutron star populations (SGRBs are much less concentrated on the light).

The choice of pixel selection radius (12\,kpc) has only a small impact on the resultant F$_\mathrm{light}$ values. For the interested reader, we have made python scripts and notebooks available which generate a similar F$_\mathrm{light}$ figure, given user defined input parameters\footnote{\url{https://github.com/achrimes2/MW-NS-Flight}}.

\begin{figure*}
\centering
\includegraphics[width=0.95\textwidth]{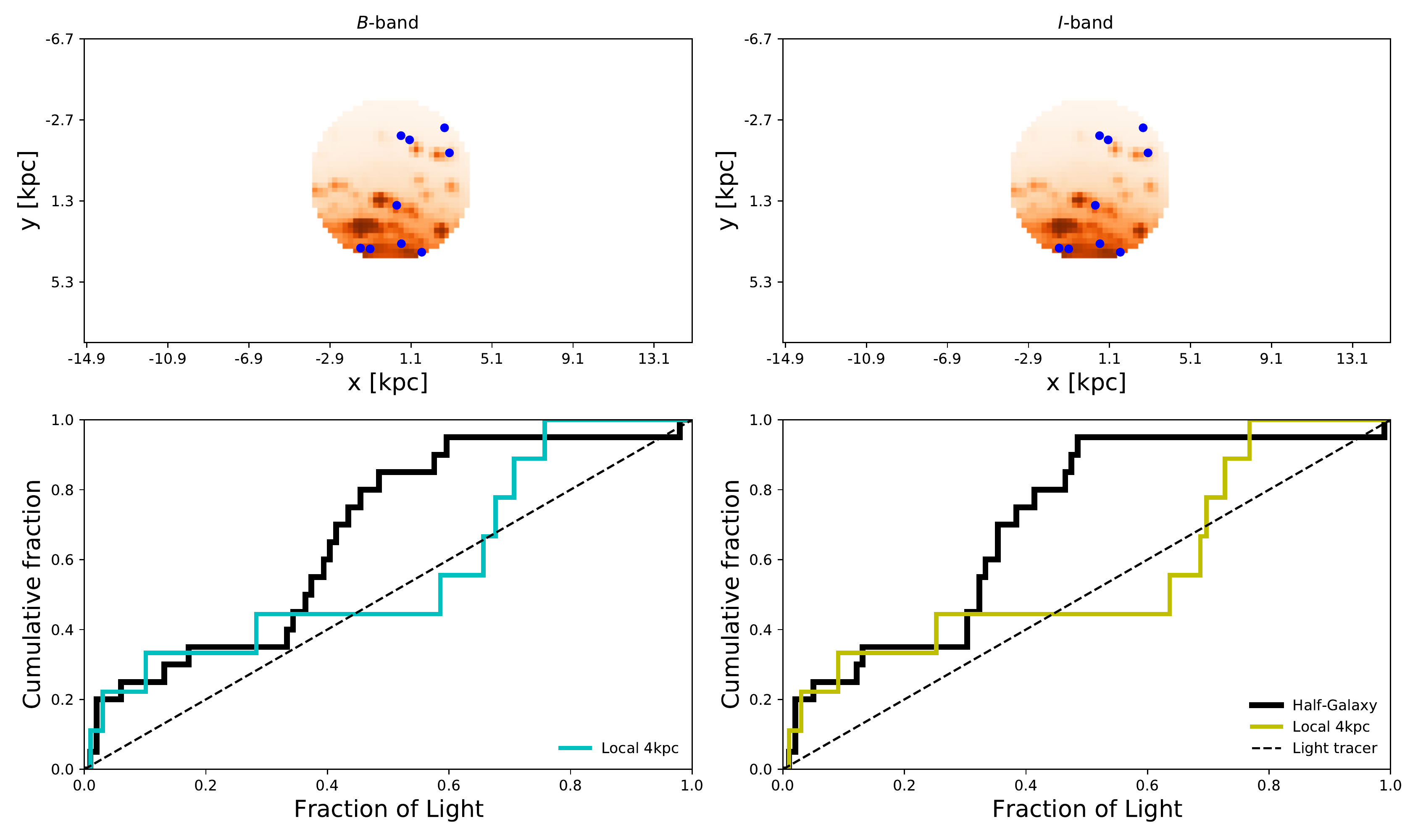}
\caption{Exploring the effect of restricting the F$_\mathrm{light}$ analysis to the local disc ($<4$\,kpc). This region has more accurate distances estimates for both the Galactic neutron stars and spiral-arm tracing masers. It also allows us to quantify the effect of the bulge on the distributions shown in Fig. \ref{fig:flightBI}, sampling a more uniform stellar population. Top: the pixel selection for F$_\mathrm{light}$ when a 4\,kpc heliocentric distance restriction is applied. Bottom: magnetar F$_\mathrm{light}$ distributions for the half Galaxy as shown in Fig. \ref{fig:flightBI}, and the local region results in each band. Excluding the bulge pushes the distribution towards the 1:1 light-tracer line.}
\label{fig:helio_magnetars}
\end{figure*}

\begin{table}
\centering 
\caption{As for Tables \ref{tab:ADoffset} and \ref{tab:ADencflux}, but for half-Galaxy $B$-band (top four rows) and $I$-band (bottom four rows) F$_\mathrm{light}$ comparisons, as shown in Figs. \ref{fig:flightBI}, \ref{fig:helio_pulsars}, \ref{fig:helio_LMXBs} and \ref{fig:helio_HMXBs}. } 
\label{tab:ADflightB}
\begin{tabular}{llllllll}
\hline %
           & LGRB                               & SLSN                               & FRB                                & SN Ia                              & SGRB \\
\hline %
Magnetars & 0.02                                  & 0.01                                  & \cellcolor{shade}\textbf{0.25} & 0.01 & 0.01 \\
HMXRBs    & 0.01                                  & 0.00                                  & \cellcolor{shade}\textbf{0.11} & 0.00 & 0.00 \\
Pulsars   & 0.00                                  & 0.00                                  & \cellcolor{shade}\textbf{0.21} & 0.00 & 0.00 \\
LMXRBs    & \cellcolor{shade}\textbf{0.25} & \cellcolor{shade}\textbf{0.20} & \cellcolor{shade}\textbf{0.25} & 0.00 & 0.00 \\
\hline %
           & LGRB                               & CCSNe                              & FRB                                & SN Ia                              & SGRB \\
\hline %
Magnetars & 0.00 & 0.00                                  & \cellcolor{shade}\textbf{0.25} & 0.03                                  & 0.00 \\
HMXRBs    & 0.00 & 0.00                                  & \cellcolor{shade}\textbf{0.25} & 0.00                                  & 0.00 \\
Pulsars   & 0.00 & 0.00                                  & \cellcolor{shade}\textbf{0.25} & 0.00                                  & 0.00 \\
LMXRBs    & 0.00 & \cellcolor{shade}\textbf{0.10} & \cellcolor{shade}\textbf{0.25} & \cellcolor{shade}\textbf{0.25} & 0.00 \\
\hline %
\end{tabular}
\end{table}

\begin{table}
\centering 
\caption{F$_\mathrm{light}$ AD-test  results for comparisons within the local 4\,kpc only, as shown in Figs. \ref{fig:helio_magnetars}, \ref{fig:helio_pulsars}, \ref{fig:helio_LMXBs} and \ref{fig:helio_HMXBs}. Top four rows: $B$-band, bottom four: $I$-band.} 
\label{tab:ADregionA}
\begin{tabular}{llllll}
\hline
           & LGRB                               & SLSN                               & FRB                                & SN Ia                              & SGRB \\
\hline
Magnetars & \cellcolor{shade}\textbf{0.25} & \cellcolor{shade}\textbf{0.25} & \cellcolor{shade}\textbf{0.22} & \cellcolor{shade}\textbf{0.19} & 0.00 \\
HMXRBs    & \cellcolor{shade}\textbf{0.25} & \cellcolor{shade}\textbf{0.24} & \cellcolor{shade}\textbf{0.25} & 0.04                                  & 0.00 \\
Pulsars   & \cellcolor{shade}\textbf{0.07} & 0.04                                  & \cellcolor{shade}\textbf{0.25} & 0.01                                  & 0.00 \\
LMXRBs    & \cellcolor{shade}\textbf{0.22} & \cellcolor{shade}\textbf{0.11} & \cellcolor{shade}\textbf{0.15} & 0.02                                  & 0.00 \\
\hline
           & LGRB                               & CCSNe                              & FRB                                & SN Ia                              & SGRB \\
\hline
Magnetars & \cellcolor{shade}\textbf{0.09} & \cellcolor{shade}\textbf{0.23} & \cellcolor{shade}\textbf{0.25} & \cellcolor{shade}\textbf{0.25} & 0.00 \\
HMXRBs    & \cellcolor{shade}\textbf{0.06} & \cellcolor{shade}\textbf{0.25} & \cellcolor{shade}\textbf{0.25} & \cellcolor{shade}\textbf{0.25} & 0.00 \\
Pulsars   & 0.00                                  & \cellcolor{shade}\textbf{0.14} & \cellcolor{shade}\textbf{0.25} & \cellcolor{shade}\textbf{0.25} & 0.00 \\
LMXRBs    & \cellcolor{shade}\textbf{0.13} & \cellcolor{shade}\textbf{0.25} & \cellcolor{shade}\textbf{0.19} & \cellcolor{shade}\textbf{0.25} & 0.00 \\
\hline %
\end{tabular}
\end{table}

\subsection{Local region results}
In Sections \ref{sec:mappingNS} and \ref{sec:mappingMW}, it was noted that sample completeness is higher, and distance estimates more reliable, in the local disc (within 3-5\,kpc of our location). This is because trigonometric parallax measurements are possible in this region (both for the neutron stars and the masers), and survey sensitivities are less of a issue. There are also other biases, such as an absence of pulsars around the Galactic centre due to the high dispersion measures here \citep[][and see Fig. \ref{fig:milkymap}]{2017MNRAS.471..730R}. Furthermore, using the half-Galaxy map mixes the older population of the bulge and the younger, higher star formation rate disc. In summary, our half-Galaxy comparisons to extragalactic samples may be suffering from biases which have not been fully corrected.

In order to quantify these effects on our results, we present F$_\mathrm{light}$ AD-test results for the local 4\,kpc region versus extragalactic transient samples. In this region, completeness should be higher and distance uncertainties lower, as a higher proportion of the sample (particularly XRBs) have parallax-based distance estimates. We select pixels and neutron stars within 4\,kpc of the Sun and $<12$\,kpc from Galactic centre, and re-calculate the F$_\mathrm{light}$ distribution for this area only. The magnetar results are shown in Fig. \ref{fig:helio_magnetars}, alongside the half-Galaxy results to demonstrate the difference. The corresponding pulsar and XRBs distributions are again shown in Figs. \ref{fig:helio_pulsars}, \ref{fig:helio_LMXBs} and \ref{fig:helio_HMXBs} of Appendix \ref{apx:B}. 

Restricting the F$_\mathrm{light}$ analysis to the local disc has a larger impact in the $I$-band than the $B$-band, but in both cases has shifted the magnetars/pulsars/XRBs closer towards being light-tracers. This makes sense in terms of the Galactic morphology: younger magnetars and HMXB systems avoid the older (but bright) bulge, so the inclusion of the bulge in F$_\mathrm{light}$ calculations pushes them away from being unbiased tracers. Similarly, LMXBs - which are concentrated in the bulge - are shifted the least by its exclusion from F$_\mathrm{light}$. This also explains why the shift between half-Galaxy and local results is greater in $I$, as the bulge contributes a higher proportion of the total light in that band. 

Table \ref{tab:ADregionA} lists the local region AD-test results. The results are less clear compared to the half-Galaxy comparisons. FRBs are again a good match to Galactic neutron stars, and the significance of this is even more apparent when the p-values are considered (reaching the 0.25 cap 6/8 times). However, unlike the half-Galaxy case, LGRBs, SLSNe, CCSNe and thermonuclear SNe are also consistent with several Galactic populations. Plausibly, the smaller Galactic sample sizes that result from restricting ourselves to $<4$\,kpc reduces the power of the AD-test, making consistency with other samples harder to rule out.

\subsection{Fraction of light uncertainties}
Aside from the impact of the multi-component nature of the Galaxy on these results, uncertainties in both neutron star distances and spiral arm structure also play a role in shaping the distributions. For example, it is possible that the neutron stars appear artificially less concentrated with respect to light due to distance uncertainties. This should preferentially scatter them away from their natal star forming regions, assuming they were born in star forming regions, given that inter-arm regions take up more volume than the arms. The local region results should address this somewhat - in addition to removing the effect of the Galactic bulge, parallax measurements are possible here (both in the optical for XRBs and radio for pulsars). 

A reason to suspect that arm position inaccuracies are not a major issue is that the spiral arm model is derived directly from trigonometric parallax measurements of CO/HI emission in the local disc \citep[][]{2016ApJ...823...77R}. CO/HI tracers have previously been demonstrated to better trace arm mid-points than masers \citep{2014AJ....148....5V} - we are simply populating the CO/HI derived spiral arm model with the masers. Furthermore, the images we construct have a spatial resolution of 250\,pc/pixel and a PSF to replicate {\it HST}-quality imaging, which is insufficient to resolve the offset between arm mid-points and typical CO/HI offsets of 100-200\,pc \citep{2014AJ....148....5V}. In the local few kiloparsecs the spiral arm tracer distances are directly measured, beyond this the arms are extrapolated, but throughout we have neglected the far side of the disc where this extrapolation would be particularly problematic. 

While the local 4\,kpc comparisons are inherently less affected by uncertainties, the full half-Galaxy results are still valuable, as many transients do occur in galaxies with mixed populations and multiple components. This is particularly notable for FRBs, whose host galaxies are varied and include several Milky Way-like spirals and barred spirals. 

The other assumptions we have made about the Galactic structure and the colour of the different components also contribute uncertainty, such as the adopted arm widths, which may have been underestimated. \citet{2021A&A...650A.112Z} show that young stars are only loosely concentrated in the spiral arms \citep[][find a similar result for HMXBs]{2021arXiv210202615A}. Widening the arms by changing Equation \ref{eq:xygaussian} in Section \ref{sec:mappingMW} would tend to increase the association of Galactic neutron stars with light. This is because any distance uncertainty induced offsets from nearby bright pixels will be reduced. However, the pixel values would also decrease (there would be fewer masers per pixel). 

Overall, specific Galaxy mapping choices appear to be less important than Galactic neutron star sample incompleteness and distance uncertainties. We refer the reader to the interactive tools at \url{https://github.com/achrimes2/MW-NS-Flight}, which can be used to vary the relative light contributions of the bulge/bar, disc and arms, to visualise the impact on F$_{\mathrm{light}}$. The spatial resolution of the image can also be degraded below the 250\,kpc per pixel level used in this paper.

\section{Discussion}\label{sec:discuss}
\subsection{Summary of results}
Table \ref{tab:results} gives an overview of our results, in terms of the percentage of times that the different extragalactic transients agree at the 2${\sigma}$ level ($p>0.05$) with Galactic neutron stars, as measured on the Milky Way face-on image. The comparisons considered are the Galactocentric offsets and host-normalised offsets in both bands, the $I$-band enclosed flux, the $B$ and $I$-band half-Galaxy F$_\mathrm{light}$ and the local 4\,kpc $B$ and $I$-band F$_\mathrm{light}$. The total number of AD-tests made against each transient is counted across Tables \ref{tab:ADoffset}, \ref{tab:ADencflux}, \ref{tab:ADflightB} and \ref{tab:ADregionA}. 

It is clear from Table \ref{tab:results} that the distribution of Galactic neutron stars on the Milky Way best matches, of all of the transients, FRBs on their hosts. However, we are unable to clearly differentiate which class of neutron star is the best fit - within the FRB AD-tests, magnetars return $p>0.05$ on 6 occasions, HMXBs/pulsars 5 times each, and LMXBs 8 times. 

Of the other extragalactic transients, SGRBs \citep[known to be mergers involving at least one neutron star,][]{2005Natur.438..994B,2010ApJ...708....9F,2011MNRAS.413.2004C,2014MNRAS.437.1495T,2017ApJ...848L..12A}, are the worst match. SRGBs have extended offset distributions, due to the combination of natal kicks and long gravitation wave inspiral times. The implication is that, in the Milky Way, old, kicked systems are missing from our catalogues, possibly because they are distant and not bright emitters like magnetars, pulsars or XRBs. Of the other samples, none stand out as being more or less consistent with the Galactic systems.

\begin{table}
\centering 
\caption{For each extragalactic transient, the number of comparisons made to magnetar, pulsar and XRB systems on the Milky Way is listed, along with the number and fraction of those tests which return at p-value$>0.05$. The results used to populate this table are listed in Tables \ref{tab:ADoffset} (offsets and host normalised offsets in both bands), \ref{tab:ADencflux} (enclosed fluxes, $I$-band only), \ref{tab:ADflightB} (half-Galaxy F$_\mathrm{light}$) and \ref{tab:ADregionA} (F$_\mathrm{light}$ in the local 4\,kpc). Overall, FRBs are clearly distributed on their hosts in a similar manner to neutron stars on the Milky Way, and are a better match than the other transients tested.} 
\label{tab:results}
\begin{tabular}{lccc}
\hline %
Transient & N$_\mathrm{AD-test} > 0.05$ & N$_\mathrm{AD-test}$ & Fraction $>0.05$\\
\hline %
LGRB & 8 & 28 & 0.29 \\
SLSN & 4 & 20 & 0.20 \\
CCSNe & 6 & 28 & 0.21 \\
FRB & 24 & 32 & 0.75 \\
SNe Ia & 8 & 24 & 0.33 \\
SGRB & 2 & 28 & 0.07 \\
\hline 
\end{tabular}
\end{table}

\subsection{Implications for the neutron star-FRB connection}
Initial FRB host population studies \citep{2020ApJ...903..152H,2020arXiv201211617M} led to differing interpretations over whether magnetar flares can explain all FRBs \citep{2020arXiv200913030B,2020ApJ...905L..30S,2020ApJ...899L..27M}. It is also unclear whether the identification of FRB-like flares from SGR\,1935 \citep{2020Natur.587...59B,2020Natur.587...54C} definitively establishes a connection between magnetars and extragalactic FRBs \citep{2021NatAs.tmp...31T,2021NatAs.tmp...30R,2021MNRAS.tmp..771B}.

Our results, looking at magnetars, pulsars and XRBs on the Milky Way's light, show that Galactic neutron stars are consistent with FRB locations on their hosts. This result holds across various types of comparison (distance to the star forming region, host offset, enclosed flux and fraction of light), although the significance varies substantially between these. Other extragalactic transients are also consistent with the Milky Way neutron star distribution, depending on the comparison made, but FRBs are the transient most frequently in agreement. Our results cannot differentiate between most FRB progenitor models which invoke neutron stars \citep[e.g. magnetars from core-collapse and accretion induced collapse, ultra luminous X-ray binaries, combing models etc,][]{2019ApJ...886..110M,2020ApJ...893L..26I,2020MNRAS.498.1397L,2021arXiv210206138S,2021ApJ...907..111Z,2021arXiv210206796D}. If FRBs do arise from neutron star systems, we disfavour the scenario where they originate from nascent magnetars born in LGRBs or SLSNe \citep[see also][]{2020ApJ...903..152H,2020arXiv201211617M}. A caveat is that the Milky Way would be an atypical LGRB or SLSN host galaxy, so the current Galactic magnetar population may not have arisen through this pathway. Furthermore, the stellar masses and star formation rates of FRB hosts are also, on average, similar to the Milky Way \citep{2020ApJ...899L...6L,2020ApJ...895L..37B,2020ApJ...903..152H}. 

It is interesting to note that the Galactic neutron star population is less consistent with CCSNe than FRBs, perhaps surprising given that CCSNe are expected to be the dominant production channel for neutron stars. This could be reflecting the offset between where supernovae occur and where neutron stars are observed. Given typical natal kick velocities \citep[e.g.][]{2005MNRAS.360..974H,2016MNRAS.461.3747B}, for young magnetars (ages ${\sim}$10$^{3}$-10$^{5}$\,yrs) this is expected to be a small distance, but it could be hundreds of parsec for XRBs. Indeed, this has been used to argue for a XRB-like origin for FRB\,180916B, which lies ${\sim}$250\,pc from a nearby star forming region in its host \citep{2021ApJ...908L..12T}. 

Alternative pathways have also been put forward, particularly for magnetar production, including merger and accretion induced collapse of white dwarfs \citep{2019ApJ...886..110M}. The recent discovery of repeating FRBs in a host galaxy dominated by a ${\sim}$\,Gyr old stellar population \citep[FRB\,20201124A,][]{2021arXiv210611993F}, and in a globular cluster \citep[FRB\,20200120E,][]{2021arXiv210511445K}, suggests that at least some FRB progenitors/magnetars have long delay times. 

Finally, we acknowledge that the FRB host sample is currently small and therefore statistical consistency with other data by means of an AD-test is more likely (or rather, it is harder to rule out). This does not diminish the results summarised in Table \ref{tab:results}, but reflects the limitations of the current data sets. Looking solely at Table \ref{tab:results}, neutron star systems appear to be a more plausible origin for FRBs than for other transients (if we were to ignore all other knowledge that we have about the nature of the other transients). To increase the significance of this study's results, we require better accuracy and precision in Galactic neutron star distance measurements, a deeper understanding of the biases affecting these samples, and a larger sample of FRB hosts. In future, it may be possible to use these methods to determine not only whether extragalactic FRBs arise from neutron stars, but which specific systems are the progenitors. The apparent difference between single and repeating bursts \citep{2021arXiv210604356P} could also be investigated in this way, both in terms of their global host properties, and the environments sampled within them.

\section{Conclusions}\label{sec:conc}
Motivated by the possibility that magnetar activity is the origin of fast radio bursts, we have created an image of the Milky Way, simulating its face-on appearance from an extragalactic distance. We then placed magnetar, pulsar and XRB populations on the image, according to their best distance estimates, and measured how these systems are distributed with respect to Galactic light in terms of Galactocentric offsets, host normalised offsets, enclosed fluxes and the fraction of light statistic. Distributions of these measurements for extragalactic transients, including fast radio bursts, are compared to the Galactic neutron star results with AD-tests. There are $\sim$20-30 AD-tests for each transient, across the range of Galactic populations and literature measurements available for each. We find that Galactic neutron stars are distributed on the Milky Way in a similar manner to FRBs on their hosts, with 75 per cent of AD-tests returning a p-value $>0.05$. FRBs also stand out as being in better agreement with the Galactic neutron star population than other extragalactic transients. These results appear robust against incompleteness, uncertainties in Galaxy modelling and distance uncertainties. We cannot distinguish whether isolated magnetars or accreting/interacting binaries containing a neutron star are the best match, but nevertheless, these results provide further support for FRB source models which invoke neutron star systems. 

To make further progress with this method for understanding FRBs, an improved understanding of distance uncertainties and incompleteness in Galactic neutron star populations is required. Larger FRB host samples will also be key, but this population will surely grow over the coming years.

\section*{Acknowledgements}
AAC is supported by the Radboud Excellence Initiative. AJL has received funding from the European Research Council (ERC) under the European Union’s Seventh Framework Programme (FP7-2007-2013) (Grant agreement No. 725246). JDL acknowledges support from a UK Research and Innovation Fellowship (MR/T020784/1).

This work has made use of {\sc ipython} \citep{2007CSE.....9c..21P}, {\sc numpy} \citep{2020arXiv200610256H}, {\sc scipy} \citep{2020NatMe..17..261V}; {\sc matplotlib} \citep{2007CSE.....9...90H} and {\sc astropy},\footnote{https://www.astropy.org} a community-developed core Python package for Astronomy \citep{astropy:2013, astropy:2018}. We have also made use of Ned Wright's cosmology calculator \citet{2006PASP..118.1711W}.

Finally, we thank the referee for their constructive feedback, which has substantially improved the clarity of this manuscript.

\section*{Data Availability}
The input masers used in this paper were selected from the RMS survey \citep{2014MNRAS.437.1791U}. Their distances were calculated using version 2 of the parallax-based BeSSel \citep{2011AN....332..461B} distance calculator \citep{2016ApJ...823...77R,2019ApJ...885..131R}, available at \url{https://bessel.vlbi-astrometry.org/node/378}. Neutron star data were obtained from the McGill catalogue \citep[for magnetars,][]{2014ApJS..212....6O}, available at \url{https://www.physics.mcgill.ca/~pulsar/magnetar/main}; the ANTF database \citep[for pulsars,][]{2005AJ....129.1993M}, available at \url{https://www.atnf.csiro.au/research/pulsar/psrcat}; \citet[][for LMXBs]{2020NewAR..8801536S} and \citet[][for HMXBs]{2019NewAR..8601546K}.

We have made interactive python scripts and notebooks available online, which generate the F$_\mathrm{light}$ results presented in this paper, but also allow the user to change various parameters. Along with images of the Galaxy in fits format, these can be accessed at \url{https://github.com/achrimes2/MW-NS-Flight}. Also available in this repository are the F$_\mathrm{light}$, offset and enclosed flux values for the magnetar, pulsar and XRB samples used in this paper. These data can also be found on the journal website as supplementary materials.



\bibliographystyle{mnras}
\bibliography{ExtragalacticMW} 




\appendix

\section{Milky Way magnetar measurements}\label{apx:A}
In Table \ref{tab:info} of this appendix, we provide a table of measurements for the 20 Milky Way magnetars which lie this side of Galactic centre ($y<8.3$\,kpc on our map). The measurements are typical of those used in extragalactic transient studies, including host offset, host normalised offset, F$_\mathrm{light}$ and the enclosed flux fraction at their radius. The heliocentric distances, their uncertainties, and the type of measurement, are also provided \citep[partially reproduced from the McGill magnetar catalogue,][]{2014ApJS..212....6O}. 

Table \ref{tab:other} contains the same offset, enclosed flux and fraction of light data for the pulsar and XRB samples. Only a subset of pulsar data is shown here, the full results for all samples are available at \url{https://github.com/achrimes2/MW-NS-Flight} or as supplementary materials on the journal website. These results can be used in future comparative studies.

\begin{table*}
\centering 
\caption{The 20 magnetars that satisfy the criteria of having (i) a distance estimate, (ii) lying at $y<8.3$\,kpc. Magnetar distance information is from the McGill catalogue \citep[][and reference therein]{2014ApJS..212....6O}. The F$_\mathrm{light}$ values are calculated using a 12\,kpc pixel selection radius and pixels from the whole Galaxy.} 
\label{tab:info}
\begin{tabular}{lllccccccc}
\hline %
Magnetar             & Dist./Assoc. & d$_{\odot}$ {[}kpc{]} & ${\delta}r$ {[}kpc{]} & ${\delta}r$/$r_{e}$ (B) & ${\delta}r$/$r_{e}$ (I) & F$_\mathrm{light}$ (B) & F$_\mathrm{light}$ (I) & F$_\mathrm{enc}$ (B) & F$_\mathrm{enc}$ (I) \\
\hline %
4U0142+61 & RC & 3.6$^{+0.4}_{+0.4}$ & 10.85 & 4.57 & 4.88 & 0.02 & 0.01 & 0.98 & 0.98 \\
SGR0418+5729 & PA & ${\sim}$2 & 9.95 & 4.20 & 4.47 & 0.07 & 0.05 & 0.95 & 0.96 \\
SGR0501+4516 & PA/SNR & ${\sim}$2 & 10.12 & 4.27 & 4.55 & 0.03 & 0.03 & 0.96 & 0.96 \\
1E1048.1-5937 & RC & 9$^{+1.7}_{-1.7}$ & 10.10 & 4.26 & 4.54 & 0.03 & 0.03 & 0.96 & 0.96 \\
1E1547.0-5408 & SNR & 4.5$^{+0.5}_{-0.5}$ & 5.04 & 2.13 & 2.27 & 0.35 & 0.31 & 0.57 & 0.64 \\
PSRJ1622-4950 & DM/SNR & 9$^{+1.4}_{-1.4}$ & 3.97 & 1.68 & 1.78 & 0.44 & 0.39 & 0.44 & 0.53 \\
CXOUJ164710.2-455216 & CA & 3.9$^{+0.7}_{-0.7}$ & 4.75 & 2.00 & 2.14 & 0.37 & 0.32 & 0.54 & 0.61 \\
1RXSJ170849.0-400910 & RC & 3.8$^{+0.5}_{-0.5}$ & 4.59 & 1.94 & 2.06 & 0.42 & 0.36 & 0.52 & 0.59 \\
SGRJ1745-2900 & HI & 8.3$^{+0.3}_{-0.3}$ & 0.10 & 0.04 & 0.04 & 0.99 & 0.99 & 0.00 & 0.00 \\
XTEJ1810-197 & HI & 3.5$^{+0.5}_{-0.4}$ & 4.81 & 2.03 & 2.16 & 0.34 & 0.31 & 0.54 & 0.61 \\
SwiftJ1818.0-1607 & DM & 4.8$^{+1.65}_{-1.65}$ & 3.77 & 1.59 & 1.69 & 0.49 & 0.47 & 0.42 & 0.50 \\
SwiftJ1822.3-1606 & HI/HII & 1.6$^{+0.3}_{-0.3}$ & 6.67 & 2.81 & 3.00 & 0.39 & 0.32 & 0.74 & 0.78 \\
SwiftJ1834.9-0846 & SNR & 4.2$^{+0.3}_{-0.3}$ & 4.65 & 1.96 & 2.09 & 0.38 & 0.34 & 0.52 & 0.59 \\
1E1841-045 & SNR & 8.5$^{+1.3}_{-1}$ & 3.96 & 1.67 & 1.78 & 0.60 & 0.49 & 0.42 & 0.50 \\
3XMMJ185246.6+003317 & HI/SNR & ${\sim}$7.1 & 4.54 & 1.92 & 2.04 & 0.41 & 0.36 & 0.49 & 0.57 \\
SGR1935+2154 & F/SNR & 6.5$^{+3.0}_{-5.3}$ & 7.20 & 3.04 & 3.24 & 0.13 & 0.13 & 0.77 & 0.81 \\
1E2259+586 & SNR/PA & 3.2$^{+0.2}_{-0.2}$ & 9.73 & 4.11 & 4.37 & 0.17 & 0.13 & 0.94 & 0.95 \\
AXJ1845.0-0258 & HI/SNR & ${\sim}$8.5 & 4.27 & 1.80 & 1.92 & 0.59 & 0.48 & 0.45 & 0.54 \\
SGR2013+34 & HII & ${\sim}$8.8 & 10.04 & 4.24 & 4.51 & 0.03 & 0.03 & 0.95 & 0.96 \\
PSRJ1846-0258 & SNR & 6$^{+1.5}_{-0.9}$ & 4.22 & 1.78 & 1.90 & 0.46 & 0.42 & 0.46 & 0.54 \\
\hline 
\end{tabular}
\newline
RC - distance from red clump stars, PA - Perseus arm association, SNR - supernova remnant association, DM - dispersion measure distance, CA - cluster association, HI - distance from HI column density, HII - HII region association, F - distance estimate from burst flux
\end{table*}

\begin{table*}
\centering 
\caption{An extract of a table containing offset, enclosed flux and F$_\mathrm{light}$ results for the pulsar sample. The full data for the pulsars, XRBs and magnetars are available as .txt files on the journal website, and at \url{https://github.com/achrimes2/MW-NS-Flight}, with the same columns and in the same format as below. The F$_\mathrm{light}$ values quoted are for half-Galaxy measurements.} 
\label{tab:other}
\begin{tabular}{ccccccc}
\hline %
${\delta}r$ {[}kpc{]} & ${\delta}r$/$r_{e}$ (B) & ${\delta}r$/$r_{e}$ (I) & F$_\mathrm{light}$ (B) & F$_\mathrm{light}$ (I) & F$_\mathrm{enc}$ (B) & F$_\mathrm{enc}$ (I) \\
\hline %
9.64 & 4.06 & 4.33 & 0.04 & 0.04 & 0.94 & 0.95 \\
15.65 & 6.59 & 7.03 & 0.00 & 0.00 & 1.00 & 1.00 \\
8.76 & 3.69 & 3.94 & 0.06 & 0.06 & 0.88 & 0.90 \\
9.26 & 3.90 & 4.16 & 0.04 & 0.04 & 0.91 & 0.93 \\
10.99 & 4.63 & 4.94 & 0.02 & 0.02 & 0.98 & 0.99 \\
\hline 
\end{tabular}
\end{table*}

\section{Fraction of light distributions for pulsars and XRBs}\label{apx:B}
\begin{figure*}
  \centering
    \includegraphics[width=0.99\columnwidth]{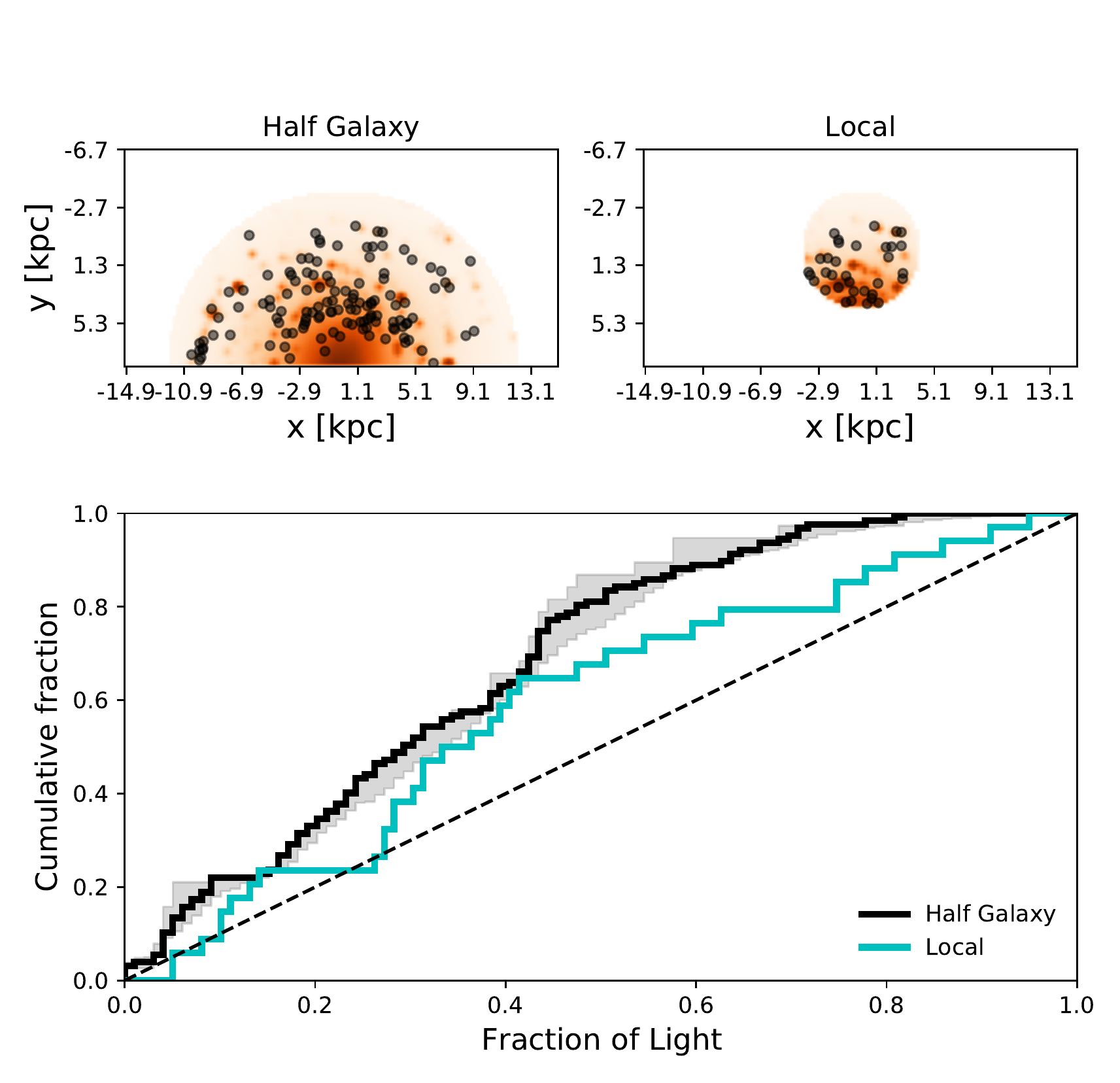}
    \includegraphics[width=0.99\columnwidth]{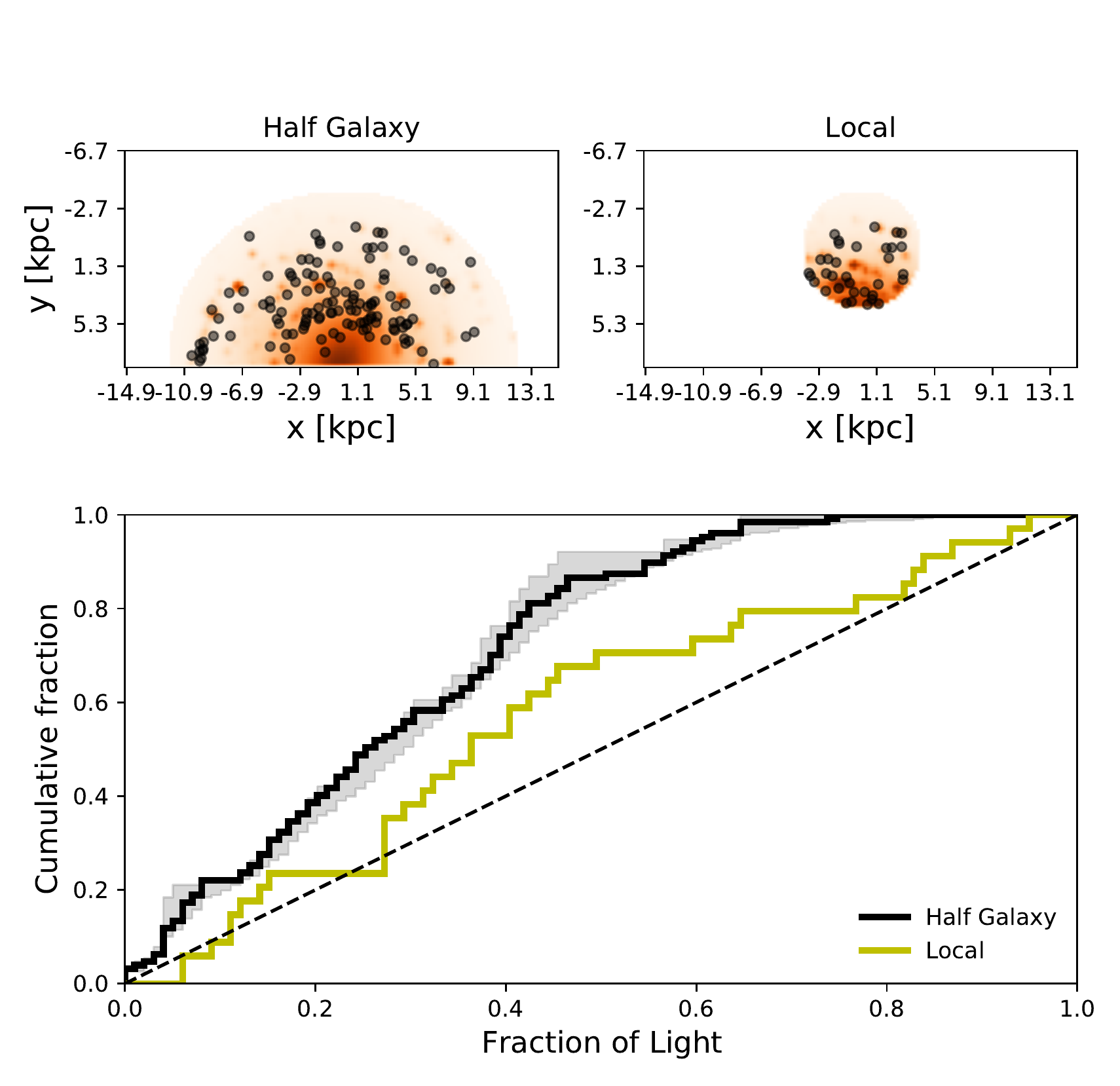}
    \caption{The same information as in Figs. \ref{fig:flightBI} and \ref{fig:helio_magnetars}, but for the bright ANTF pulsars. Top: F$_\mathrm{light}$ $B$-band and $I$-band colour maps with the half-Galaxy and 4\,kpc heliocentric distance restriction. The pulsar locations are indicated. Bottom:  the corresponding F$_\mathrm{light}$ distributions. The grey shading bounds the minimum and maximum distributions that arise from varying the luminosity cut by $\pm$0.5\,dex (see Fig. \ref{fig:pulsar_lum}). }
    \label{fig:helio_pulsars}
\end{figure*}

\begin{figure*}
  \centering
    \includegraphics[width=0.99\columnwidth]{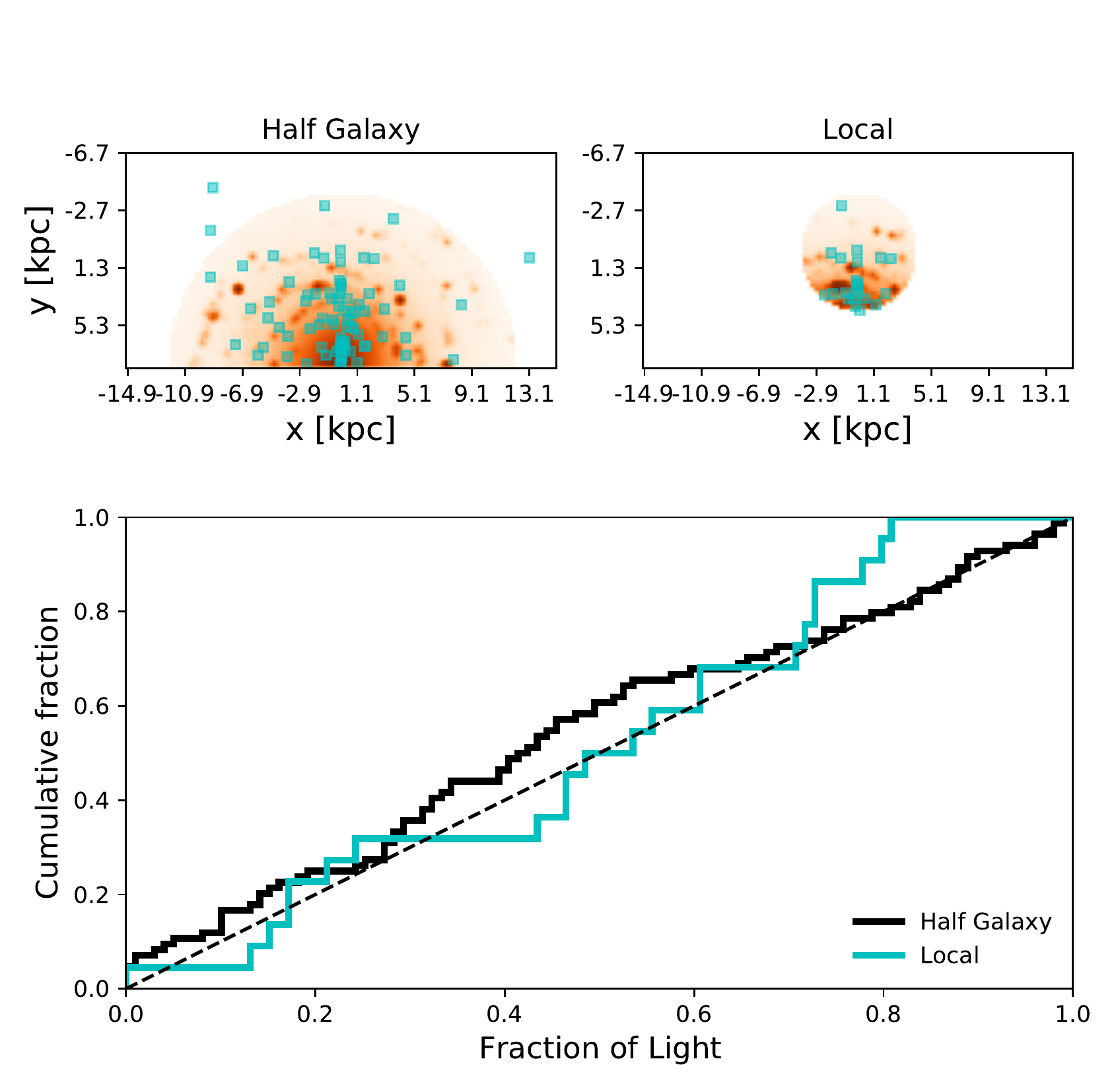}
    \includegraphics[width=0.99\columnwidth]{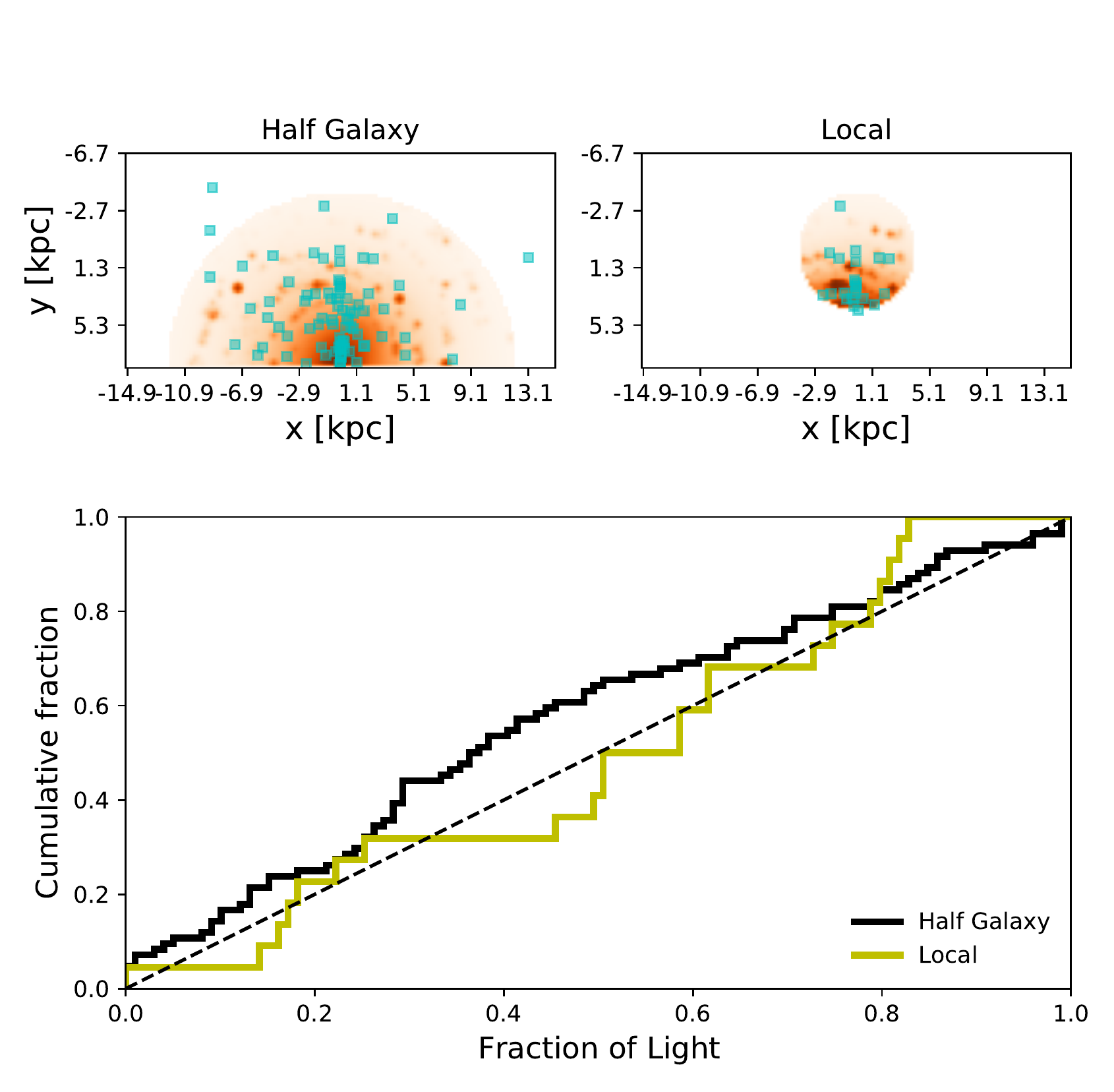}
    \caption{As in Fig. \ref{fig:helio_pulsars}, but for the LMXBs. Only in the half-Galaxy case do we consider objects outside the selected pixels, assigning them F$_\mathrm{light}=0$.}
    \label{fig:helio_LMXBs}
\end{figure*}

\begin{figure*}
  \centering
    \includegraphics[width=0.99\columnwidth]{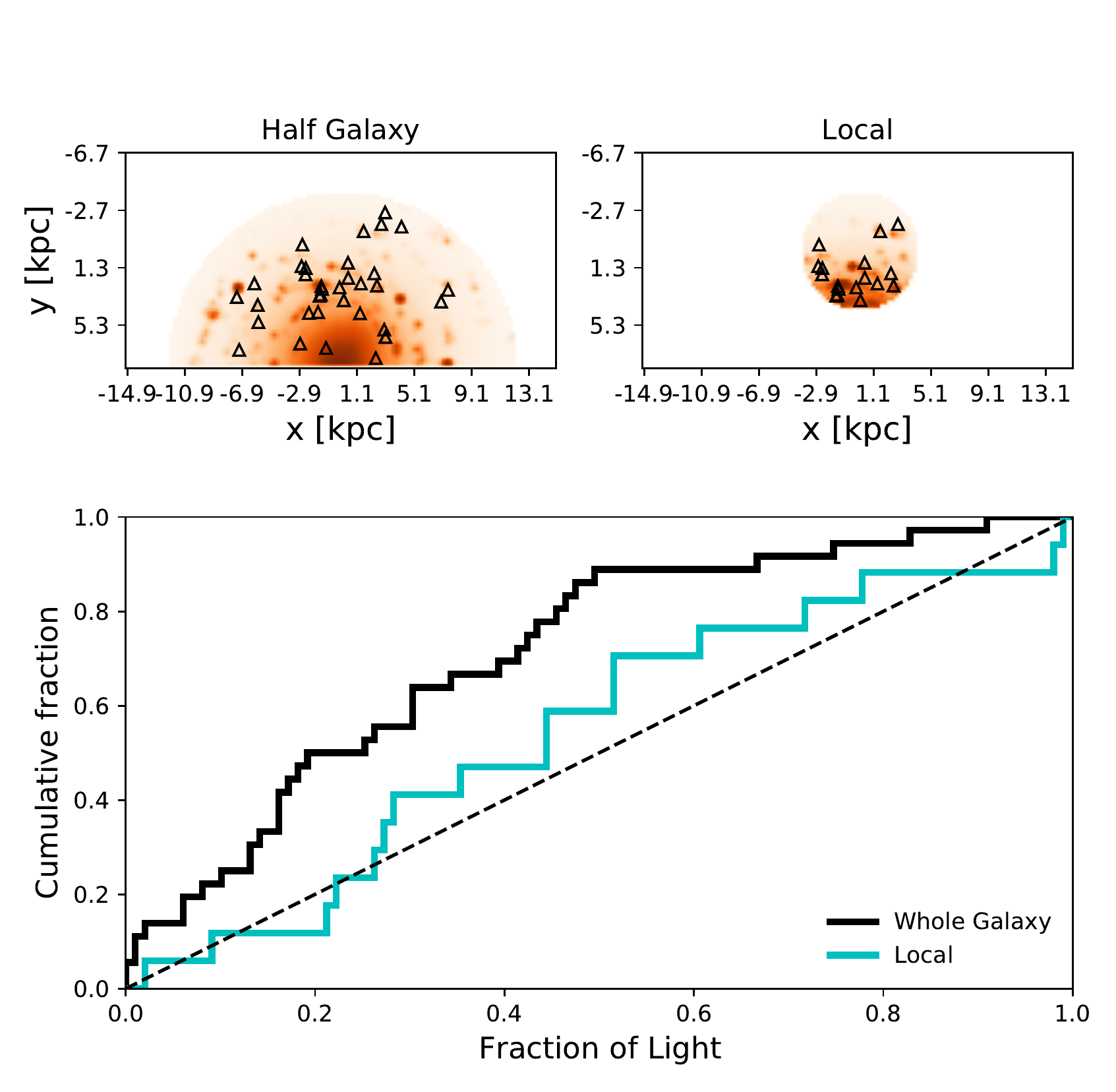}
    \includegraphics[width=0.99\columnwidth]{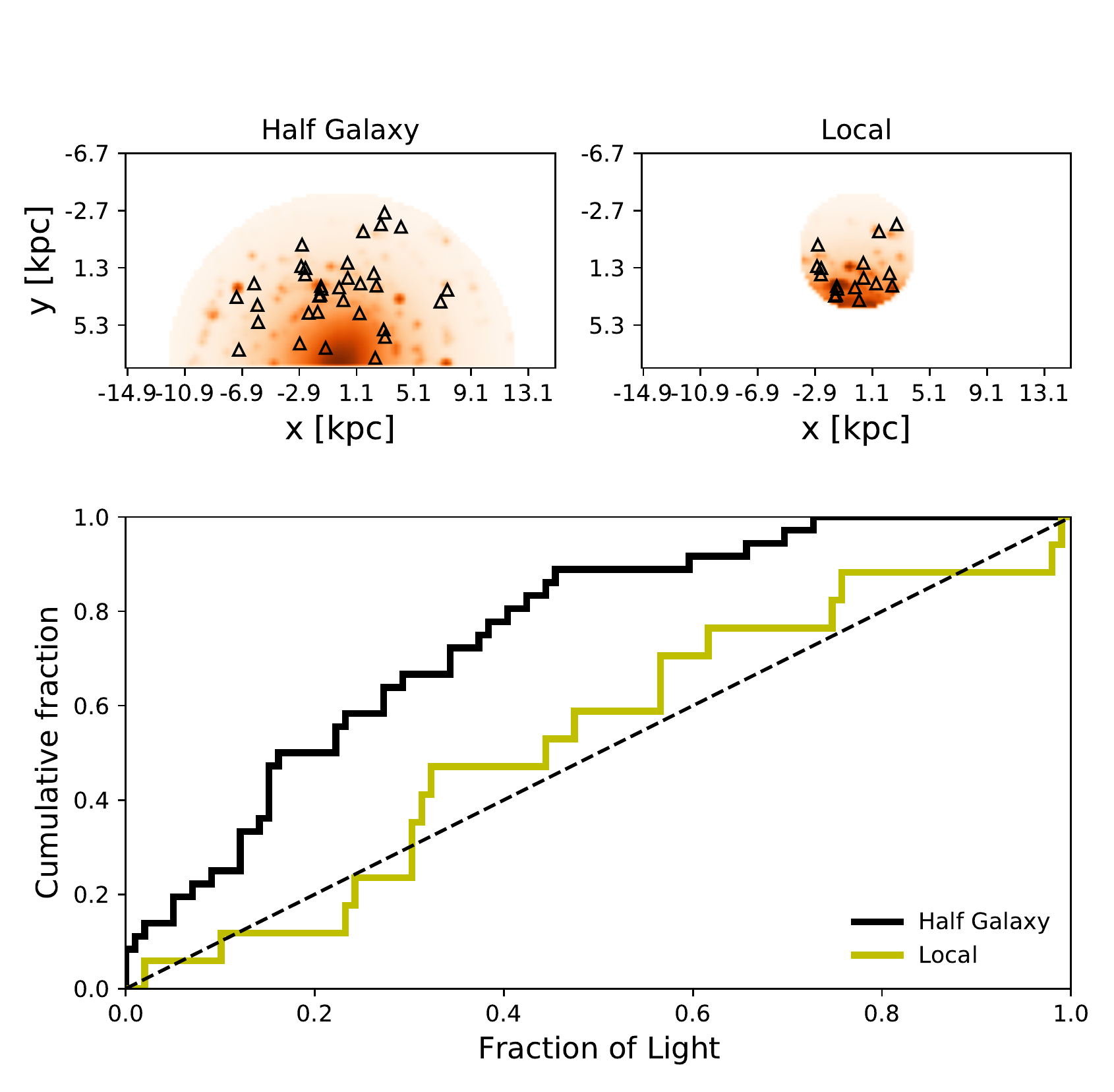}
    \caption{As in Figs. \ref{fig:helio_pulsars} and \ref{fig:helio_LMXBs}, but for HMXBs.}
    \label{fig:helio_HMXBs}
\end{figure*}

The fraction of light method described for the magnetars in Section \ref{sec:flight} is repeated here for the ATNF luminous pulsars and the INTEGRAL XRBs. Their positions on the half-map, the local sub-region, and the resultant F$_\mathrm{light}$ distributions, are shown in Figs. \ref{fig:helio_pulsars},  \ref{fig:helio_LMXBs} and \ref{fig:helio_HMXBs}. 

In Fig. \ref{fig:helio_pulsars}, the result of varying the pulsar luminosity cut shown in Fig. \ref{fig:pulsar_lum} is also demonstrated. Moving the cut within the region denoted by the dashed lines in Fig. \ref{fig:pulsar_lum}, the maximum and minimum distributions that occur in this range are shaded between, demonstrating that the precise choice of cut has a minimal impact.

As for the magnetars, restricting the analysis to the local 4\,kpc pushes the pulsar and HMXB distributions closer towards the 1:1 line, with the effect greater in the $I$-band than the $B$-band. Again, this is because these samples avoid the Galactic bulge (with an additional effect due to survey/detection biases in the pulsar case). The effect occurs to a much lesser extent for LMXBs, which being an older population, already favour the Galactic bulge.

\bsp	
\label{lastpage}
\end{document}